\documentclass{JHEP3}
\usepackage{epsfig,amsmath,amssymb}

\usepackage{rotate}

\def\half{\frac{1}{2}}
\def\third{\frac{1}{3}}

\def\lN{l_{\rm n}}
\def\lF{l_{\rm f}}

\def\low{{\rm low}}
\def\high{{\rm high}}

\def\mqlLow{m_{ql(\low)}}
\def\mqlHigh{m_{ql(\high)}}

\def\cHigh{{c2(\high)}}
\def\cLow{{c2(\low)}}

\def\mcHigh{m_{\cHigh}}
\def\mcLow{m_{\cLow}}

\def\mqlLow{m_{ql(\low)}}
\def\mqlHigh{m_{ql(\high)}}

\def\ssp{\vspace{0.6cm}}

\allowdisplaybreaks

\title{Invariant mass distributions in cascade decays}
\author{D.~J.~Miller\\ 
Department of Physics and Astronomy, 
University of Glasgow, 
Glasgow G12 8QQ, 
U.K. 
E-mail: \email{D.Miller@physics.gla.ac.uk}}
\author{P.~Osland\\ 
Department of Physics, 
University of Bergen, N-5007 Bergen, Norway\\ 
and TH Division, Physics Department, CERN, CH~1211 Gen\`eve, Switzerland \\
E-mail: \email{Per.Osland@ift.uib.no}}
\author{A.~R.~Raklev\\ 
Department of Physics, 
University of Bergen, N-5007 Bergen, Norway\\ 
and TH Division, Physics Department, CERN, CH~1211 Gen\`eve, Switzerland \\
E-mail: \email{Are.Raklev@ift.uib.no}}

\abstract{We derive analytical expressions for the shape of the invariant mass
distributions of massless Standard Model endproducts in cascade decays
involving massive New Physics (NP) particles, $D\rightarrow Cc \rightarrow Bbc
\rightarrow Aabc$, where the final NP particle $A$ in the cascade is
unobserved and where two of the particles $a$, $b$, $c$ may be
indistinguishable. Knowledge of these expressions can improve the
determination of NP parameters at the LHC. The shape formulas are composite,
but contain nothing more complicated than logarithms of simple expressions. We
study the effects of cuts, final state radiation and detector effects on the
distributions through Monte Carlo simulations, using a supersymmetric model as
an example. We also consider how one can deal with the width of NP particles
and with combinatorics from the misidentification of final state
particles. The possible mismeasurements of NP masses through ``feet'' in the
distributions are discussed.  Finally, we demonstrate how the effects of
different spin configurations can be included in the distributions.}
\keywords{SUSY, BSM, MSSM} \preprint{CERN-PH-TH/2005-121}

\begin{document}

\section{Introduction}\label{sect:intro}
While the discovery of a broken TeV-scale supersymmetry
\cite{Wess:1974tw,Fayet:1976cr,Dimopoulos:1981zb,Nilles:1983ge,Haber:1984rc}
at the LHC would solve many problems in modern particle physics, it
would also raise many questions. For example, do the forces unify at
some high scale, as the supersymmetric evolution of couplings imply;
what is the mechanism of supersymmetry breaking; does supersymmetry
provide a viable dark matter candidate?  To answer any of these
questions it is important to have accurate measurements of the
supersymmetric partner masses.

In the Minimal Supersymmetric Standard Model (MSSM), superpartner
production cross-sections at the LHC are rather large, allowing the
discovery of squarks and gluinos with masses up to about 2.5
TeV. Nevertheless, if R-parity is conserved, the traditional method of
measuring masses from resonant peaks in invariant mass distributions
cannot be easily used for superpartners. The end products of every
superpartner decay necessarily include the lightest supersymmetric
particle (LSP) which is stable and escapes the detector, preventing
full reconstruction. However, invariant mass distributions constructed
from the {\it visible} particles in a decay chain clearly do depend on
the masses of their parents, and it should be possible to extract
these masses in a systematic way. In particular, the kinematic
endpoints of these invariant mass distributions (i.e.\ their minimum
or maximum values) can be measured and their relation to the unknown
superpartner masses can be exploited. This method has been widely
studied in refs.~\cite{Hinchliffe:1996iu,Hinchliffe:1998zj,
Bachacou:1999zb,Polesello,Allanach:2000kt,Lester,Gjelsten:2004,
Gjelsten:2005aw}.

This study builds upon the work of ref.~\cite{Gjelsten:2004} which
investigated the measurement of superpartner masses via the endpoint method
for the decay\footnote{The subscripts $n$ and $f$ on the leptons, representing
``near'' and ``far'', denote the first and second lepton emitted in the
decay.}
\begin{equation}
\label{eq:squarkchain}
\tilde q_L \to \tilde \chi_2^0 q \to \tilde l_R \lN q 
\to \tilde \chi_1^0 \lF \lN q
\end{equation}
at the Snowmass mSUGRA benchmark point SPS 1a
\cite{Allanach:2002nj}. Several problems with the endpoint method were
uncovered. These include the possibility of multiple minima in global
least squares fits to the masses and mismeasurements of the endpoints
due to the occurence of ``feet'' in the distributions.

Supersymmetry is not the only New Physics (NP) scenario that can be
discovered at the LHC and which gives rise to long decay chains. In
universal extra dimension models (UED) \cite{Appelquist:2000nn},
Kaluza-Klein (KK) excitations of Standard Model (SM) particles could
have experimental signatures very similar to SUSY models. In
\cite{Cheng:2002ab} the possibility of the decay chain
\begin{equation}
\label{eq:KKchain}
Q_1 \rightarrow Z_1 q \rightarrow L_1 \lN q
\rightarrow \gamma_1 \lF \lN q
\end{equation}
was pointed out. Here the subscript 1 denotes the first KK-excitations
of the SM particles. Indeed the occurence of a similar decay chain in
any multi-particle NP model which has a good dark matter candidate
should not be a surprise. The dark matter candidate should be a weakly
interacting, neutral and massive particle, thus escaping the
detector. This particle must be kept stable by some symmetry, in our
examples R-parity for SUSY and KK-parity for UED, and this symmetry is
likely also to conserve the number of NP particles in decays, creating
a decay cascade through the hierarchy of NP masses.

In light of this we will maintain a degree of generality in our
derivation. We derive analytic formulae for the invariant mass
distributions of particles in the decay chain
\begin{equation}
D\rightarrow Cc \rightarrow Bbc \rightarrow Aabc,
\label{eq:chain}
\end{equation}
illustrated in figure~\ref{fig:cascade-chain},
\FIGURE[ht]{ \epsfxsize 12cm \epsfbox{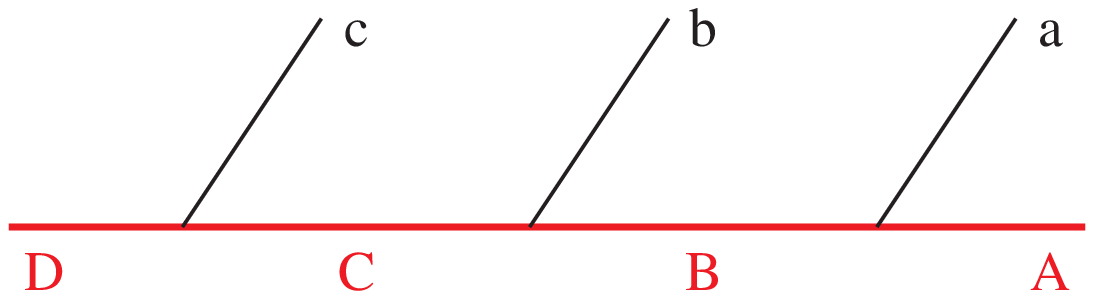}
\caption{Decay chain. Particles $c$, $b$ and $a$ are assumed massless.
\label{fig:cascade-chain}}
}
where $D$, $C$, $B$ and $A$ are massive, satisfying
\begin{equation}
m_D>m_C>m_B>m_A,
\label{eq:ordering}
\end{equation}
and the particles $a$, $b$, $c$ are taken to be massless. 

With $A$ invisible, we can form four invariant mass combinations of the decay
products $c$, $b$ and $a$: $m_{ba}$, $m_{cb}$, $m_{ca}$ and
$m_{cba}$. However, both the decay chains (\ref{eq:squarkchain}) and
(\ref{eq:KKchain}) give two opposite sign, same flavour leptons, $\lN$ and
$\lF$, which are in principle indistinguishable. This is a general feature of
decay chains; given a SM decay product with a particular flavour this flavour
should be carried on in the decay chain and may result in two same flavour
decay products. We will deal with this by calculating the shape of the
distributions of the invariant masses $m_{c2(\text{low})}$ and
$m_{c2(\text{high})}$, which are the invariant masses of $c$ combined with
either $b$ or $a$, depending on which gives the lower and higher invariant
mass.

In the following section we derive the shapes of these invariant mass
distributions. Using the decay chain (\ref{eq:squarkchain}) as an
example we consider in section~\ref{sect:parton} some complicating
effects present already at the parton level, such as final state
radiation, cuts and the width of the SUSY particles. In
section~\ref{sect:detector} we further look at detector induced
effects on the distributions and in section~\ref{sect:feet} we discuss
a possible application of the shapes of the invariant masses to
alleviate the problem with ``feet''.  In section~\ref{sect:conclusion}
we give a brief summary of our results. The inclusion of spin effects
is discussed in an appendix. An application of the shape formulas to
Monte Carlo data for the purposes of improving mass determination at
the LHC will be presented elsewhere.  
\section{Invariant mass distributions} \label{sect:IM_dist}

Our aim here is to obtain expressions for the invariant mass
distributions of cascade decays such as (\ref{eq:chain}), which can be
compared to data to fit NP particle masses. These distributions can be
easily calculated numerically by Monte Carlo integration, and indeed
we use the Monte Carlo event generator PYTHIA \cite{PYTHIA} to
numerically check our analytic results. However, Monte Carlo
integration (via phase space generation) is slow in comparison to the
evaluation of an analytic expression, which makes it impractical for
fitting masses, or performing other comparisons, where many
distributions for different masses must be generated. For this reason
we derive analytic expressions.

In the main part of the paper, we will assume that there are no spin
correlations between the subsequent decays, and thus treat the decaying
particles as if they were scalars (see, however, the appendix). Indeed,
PYTHIA~6.208 uses this simplification in its generation of phase space. This
has been shown to be a reasonable simplification for the decay chains of
physical interest as long as one does not measure the charges of the final
state particles \cite{Richardson:2001df,Barr:2004ze}, which for jets would be
very difficult to do. However, it has been suggested in
\cite{Barr:2004ze,Smillie:2005ar} that measurements of NP particle spin is
possible at the LHC because of the asymmetry between produced squarks and
antisquarks. The method used in the derivation of the invariant mass
distributions can fairly easily be adapted to a particular spin configuration
of the particles in the decay chain and distributions including spin effects
are given in the appendix for our SUSY example. For simplicity of the analytic
expressions, we will also ignore particle widths. The effects of widths can be
added later by a smearing of the distributions.

Our philosophy for calculating the invariant mass distributions will be to
write the square of the invariant masses in terms of angular variables in the
rest frame of decaying particles. In the spin-0 case these have isotropic,
flat distributions. After performing a transformation from these `flat'
variables to a set of variables which include the invariant mass under
discussion, an integration over the extra variables can be performed to
provide the desired distribution. The difficulty of the calculation will be
seen not to lie in performing the integrals, but in finding the correct
integration interval for all possible configurations of masses that obey
eq.~(\ref{eq:ordering}). Spin effects only complicate the integrand and not
the endpoints of the integration.

Distributions formed from two ``neighbouring'' particles, $m_{cb}$ and
$m_{ba}$, have simple triangular shapes when widths and spin are
ignored (see e.g.\ figure~10 and eq.~(4.2) of \cite{Gjelsten:2004}),
and will not be discussed further here. We will also limit our
discussion to cases where the intermediate NP particles $C$ and $B$
are on-shell. For a discussion on the shape of distributions in the
SUSY scenario of eq.~(\ref{eq:squarkchain}), with virtual particles in
the decay chain, see \cite{Birkedal:2005cm}.

\subsection{The two-particle invariant mass $m_{ca}$}

The invariant mass $m_{ca}$ is given by
\begin{equation}
m_{ca}^{2} \equiv \left(p_c + p_a \right)^{2} = 2 E_c^{(B)} E_a^{(B)}
\left(1-\cos\theta_{ca}^{(B)}\right),
\end{equation}
where $E_a^{(B)}$ and $E_c^{(B)}$ are the energies of $a$ and $c$
respectively and $\theta_{ca}^{(B)}$ is the angle between $c$ and $a$,
all in the rest frame of $B$, denoted by the superscript $(B)$. 
Energy and momentum conservation in the decays of the $B$, $C$ and $D$
additionally provide
\begin{eqnarray}
E_a^{(B)}&=&\frac{m_{B}^{2}-m_{A}^{2}}{2m_{B}}, 
\label{eq:EaB} \\
E_b^{(B)}&=&\frac{m_C^2-m_B^2}{2m_B}, 
\label{eq:EbB} \\
E_c^{(B)}&=&
\frac{\left(m_{D}^{2}-m_{C}^{2}\right)m_{B}}
{\left(m_{C}^{2}+m_{B}^{2}\right)-\left(m_{C}^{2}-m_{B}^{2}\right)\cos
\theta_{cb}^{(B)}},
\label{eq:EcB}
\end{eqnarray}
where $\theta_{cb}^{(B)}$ is the angle between $c$ and $b$ in
$(B)$. It will be convenient to introduce three further quantities,
\begin{align}
(m_{ca}^{\max})^{2} 
&= \frac{\left(m_{D}^{2}-m_{C}^{2}\right)
\left(m_{B}^{2}-m_{A}^{2}\right)}{m_{B}^{2}}, 
\label{eq:m_ca_max} \\
(m_{cb}^{\max})^{2} 
&= \frac{\left(m_{D}^{2}-m_{C}^{2}\right)
\left(m_{C}^{2}-m_{B}^{2}\right)}{m_{C}^{2}},
\label{eq:m_cb_max} \\
(m_{c2{\rm (eq)}}^{\max})^{2}
&=\frac{\left(m_{D}^{2}-m_{C}^{2}\right)
\left(m_{B}^{2}-m_{A}^{2}\right)}{2m_{B}^{2}-m_{A}^{2}},
\label{eq:m_c_eq}
\end{align}
where $m_{ca}^{\max}$ and $m_{cb}^{\max}$ are the maximium
possible values of $m_{ca}$ and $m_{cb}$ respectively, while
$m_{c2{\rm (eq)}}$ is a possible maximum of the $m_\cLow$
distribution (see sect.~\ref{sect:cLow}).

\ssp

From the above, it follows that
\begin{equation}
m_{ca}^{2} = (m_{ca}^{\max})^{2}
\frac{m_{B}^{2}\left(1-\cos\theta_{ca}^{(B)}\right)}
{\left(m_{C}^{2}+m_{B}^{2}\right)-\left(m_{C}^{2}-m_{B}^{2}\right)
\cos\theta_{cb}^{(B)}}.
\label{eq:m_ca}
\end{equation}
Following our philosophy of writing the invariant mass in terms of
variables with flat distributions we now express
$\cos\theta_{cb}^{(B)}$ in terms of $\cos\theta_{cb}^{(C)}$, the same
angle in the rest frame of $C$. If $C$ is a scalar this angle will be
isotropically distributed. If $C$ is {\it not} a scalar, the angular
distribution will depend on the helicities of $b$ and $c$. However,
summing over particles and antiparticles in the final state will
cancel these spin correlations and return an isotropic
distribution. This latter case is applicable here.

In the rest frame of $C$ we have the familiar result
\begin{equation}
m_{cb}^{2}=(m_{cb}^{\max})^{2}
\left(\frac{1-\cos\theta_{cb}^{(C)}}{2}\right).
\label{eq:m_cb}
\end{equation}
Using eqs.~(\ref{eq:EbB}) and (\ref{eq:EcB}), the same invariant mass in
$(B)$ is given by
\begin{equation}
m_{cb}^{2}
=(m_{cb}^{\max})^{2}
\frac{m_{C}^{2}\left(1-\cos\theta_{cb}^{(B)}\right)}
{\left(m_{C}^{2}+m_{B}^{2}\right)-\left(m_{C}^{2}-m_{B}^{2}\right)
\cos\theta_{cb}^{(B)}}.
\end{equation}
Equating the two expressions and solving for $1-\cos \theta_{cb}^{(B)}$ gives 
\begin{equation}
1-\cos\theta_{cb}^{(B)}=2\biggl[1+\frac{m_{C}^{2}}{m_{B}^{2}}
\frac{1+\cos\theta_{cb}^{(C)}}{1-\cos\theta_{cb}^{(C)}}\biggr]^{-1},
\label{eq:costheta_cb}
\end{equation}
and consequently
\begin{equation}
m_{ca}^{2} 
= (m_{ca}^{\max})^{2}
\left[\frac{m_{B}^{2}}{m_{C}^{2}}
\left(\frac{1-\cos\theta_{cb}^{(C)}}{2}\right)
+\left(\frac{1+\cos\theta_{cb}^{(C)}}{2}\right)\right]
\left(\frac{1-\cos\theta_{ca}^{(B)}}{2}\right).
\end{equation}

\ssp

We have now written the invariant mass in terms of quantities which
have isotropic distributions. Adopting the notation
\begin{equation}
u \equiv \frac{1-\cos\theta_{cb}^{(C)}}{2}, \qquad
v \equiv \frac{1-\cos\theta_{ca}^{(B)}}{2}
\label{eq:indep-var}
\end{equation}
we observe that the differential decay widths for these observables
are flat for $0 \le (u,\,v) \le 1$:
\begin{equation}
\frac{1}{\Gamma_0} \frac{\partial^2 \Gamma_0}{\partial u \partial v} 
= \theta(1-u) \theta(u) \theta(1-v) \theta(v),
\label{eq:d2Gdudv}
\end{equation}
where $\theta(x)$ is the usual step function, and where the subscript ``0''
is a reminder that spin correlations are omitted (however, see the Appendix).
We can now make a change of variables from $(u,v)$ to $(u,m_{ca}^2)$ with 
\begin{equation}
m_{ca}^{2}
=(m_{ca}^{\max})^{2}\left(1-au\right)v,
\label{eq:m_ca-other}
\end{equation}
where, for abbreviation of the equations to follow, we have written
\begin{equation}
\label{eq:def-a}
a \equiv 1-\frac{m_{B}^{2}}{m_{C}^{2}}, \quad
0 < a < 1.
\end{equation}
This gives the differential distribution
\begin{eqnarray}
\frac{1}{\Gamma_0} 
\frac{\partial^2 \Gamma_0}{\partial u \partial m_{ca}^2}
&=&\left| \frac{\partial (u,v)}{\partial (u,m_{ca}^2)} \right|
\theta(1-u) \theta(u) \theta(1-v) \theta(v) \nonumber \\
&=&
\hat \theta \left( \frac{m_{ca}^2}{(m_{ca}^{\max})^2 (1-au)}\right)  
\frac{\hat \theta(u)}{(m_{ca}^{\max})^2 (1-au)},
\end{eqnarray}
where for ease of notation we have defined a ``top-hat'' function
$\hat \theta(x) \equiv \theta(x) \theta(1-x)$.  Finally we must now
integrate over $u$ to find the distribution of $m_{ca}^{2}$:
\begin{eqnarray}
\frac{1}{\Gamma_0} \frac{\partial \Gamma_0}{\partial m_{ca}^2} 
&=&
\int_{-\infty}^{\infty} \frac{1}{\Gamma_0} \frac{\partial^2
\Gamma_0}{\partial u \partial m_{ca}^2} du \nonumber \\ 
&=& 
\int_{0}^{1}
\hat \theta
\left( \frac{m_{ca}^2}{(m_{ca}^{\max})^2 (1-au)}\right)
\frac{1}{(m_{ca}^{\max})^2 (1-au)} du \nonumber \\ &=&
\int_{0}^{u_{\rm max}} \frac{1}{(m_{ca}^{\max})^2 (1-au)} du
\end{eqnarray}
for $0 \le m_{ca} \le m_{ca}^{\max}$, where
\begin{equation}
u_{\max} = \min \left(1, \frac{1}{a}
\left[1- \frac{m_{ca}^2}{(m_{ca}^{\max})^2} \right] \right).
\end{equation}
Evaluating the integrals we find
\begin{equation}
\frac{1}{\Gamma_0} \frac{\partial \Gamma_0}{\partial m_{ca}^2} 
= \left\{ \begin{array}{lcl}
\displaystyle
\frac{1}{(m_{ca}^{\max})^{2}a}
\ln\frac{m_{C}^2}{m_{B}^2}
&\; {\rm for }\;&  
\displaystyle 0<m_{ca}<\frac{m_{B}}{m_{C}} m_{ca}^{\max},\\[5mm]
\displaystyle 
\frac{1}{(m_{ca}^{\max})^{2}a}
\ln\frac{(m_{ca}^{\max})^2}{m_{ca}^2}
&\; {\rm for }\;&  
\displaystyle \frac{m_{B}}{m_{C}} m_{ca}^{\max}
<m_{ca}<m_{ca}^{\max},
\end{array} \right.
\label{eq:mca}
\end{equation}
and zero otherwise.

From the distribution of the square of an invariant mass $m^2$, it is
a trivial task to find the distribution of the invariant mass as
\begin{equation}
\frac{1}{\Gamma_0}\frac{\partial \Gamma_0}{\partial m}
=2m\frac{1}{\Gamma_0}\frac{\partial \Gamma_0}{\partial m^2}.
\label{eq:m2tom}
\end{equation}
%
\subsection{The two-particle invariant mass $m_\cHigh$}
\label{sect:cHigh}

Having demonstrated the basic method for the simplest non-trivial case,
we now turn to the more complicated problem of finding the
distribution of the observable quantity $m_\cHigh$, defined by
\begin{equation}
m_\cHigh \equiv \max\left(m_{cb},m_{ca}\right).
\end{equation}
Here, $m_{cb}^2$ and $m_{ca}^2$ are given by eqs.~(\ref{eq:m_cb}) and
(\ref{eq:m_ca-other}), respectively. In the notation of
eq.~(\ref{eq:indep-var}), the required invariant mass squared can be written
as
\begin{equation}
m_\cHigh^2
=\max\left[(m_{cb}^{\max})^{2}u,
(m_{ca}^{\max})^{2}\left(1-au\right)v\right].
\label{eq:mc2High}
\end{equation}
We now introduce a new variable,
\begin{equation}
x=(m_{cb}^{\max})^{2}u
-(m_{ca}^{\max})^{2}\left(1-au\right)v,
\label{eq:def-x}
\end{equation}
so that the sign of $x$ picks out which of $m_{cb}$ or $m_{ca}$ is larger. Then
\begin{equation}
m_\cHigh^2=\theta(x)(m_{cb}^{\max})^{2}u \nonumber \\
+\theta(-x)(m_{ca}^{\max})^{2}\left(1-au\right)v.
\end{equation}
The new variables can be inverted to give
\begin{equation}
u=\frac{m_\cHigh^{2}+\theta(-x)x}{(m_{cb}^{\max})^{2}},\qquad
v=\frac{m_\cHigh^{2}-\theta\left(x\right)x}
{(m_{ca}^{\max})^{2}\left(1-a\frac{m_\cHigh^{2}+\theta(-x)x}
{(m_{cb}^{\max})^{2}}\right)}.
\end{equation}
Since the double-differential width with respect to $u$ and $v$ is flat
[see eq.~(\ref{eq:d2Gdudv})], we may write
\begin{eqnarray}
\frac{1}{\Gamma_0} \frac{\partial^2 \Gamma_0}{\partial x \partial m_\cHigh^2}
&=&\left| \frac{\partial (u,v)}{\partial (x,m_\cHigh^2)} \right|
\hat \theta(u) \hat \theta(v)\nonumber \\
&=& \frac{\hat \theta(u) \hat \theta(v)}
{(m_{ca}^{\max})^2 (m_{cb}^{\max})^2 (1-au)},
\label{eq:d2Gdxdmhigh}
\end{eqnarray}
and integrate over $x$ to give the desired distribution:
\begin{eqnarray}
\label{Eq:dist-c2-high}
\frac{1}{\Gamma_0} \frac{\partial \Gamma_0}{\partial m_\cHigh^2}
&=& \phantom{+}
\int_0^{\infty} 
\hat \theta \left( u_+ \right)
\hat \theta \left( v_+ \right) 
\frac{1}{(m_{ca}^{\max})^2 (m_{cb}^{\max})^2 (1-au_+)} dx
\nonumber \\
&&+ 
\int_{-\infty}^0
\hat \theta \left( u_- \right)
\hat \theta \left( v_- \right) 
\frac{1}{(m_{ca}^{\max})^2 (m_{cb}^{\max})^2 (1-au_-)} dx.
\end{eqnarray}
In the above we have written $u_{\pm}$ and $v_{\pm}$ for $u$ and $v$ in the
case of positive/negative $x$, i.e.,
\begin{eqnarray}
\lefteqn{u_- = \frac{m_\cHigh^{2}+x}{(m_{cb}^{\max})^{2}},} \hspace{6.5cm} &&
u_+ = \frac{m_\cHigh^{2}}{(m_{cb}^{\max})^{2}},
\label{eq:upm} \\
\lefteqn{v_- = \frac{m_\cHigh^{2}}
{(m_{ca}^{\max})^{2}\left(1-a\frac{m_\cHigh^{2}+x}
{(m_{cb}^{\max})^{2}}\right)},} \hspace{6.5cm} &&
v_+ = \frac{m_\cHigh^{2}-x}
{(m_{ca}^{\max})^{2} \left(1-a\frac{m_\cHigh^{2}}
{(m_{cb}^{\max})^{2}}\right)}.
\label{eq:vpm}
\end{eqnarray}
While the integrals themselves are trivial, we must take special care
with the integration endpoints. The step-functions restrict
$0<u_\pm<1$ and $0<v_\pm<1$, which in turn give restrictions on $x$
and/or $m_\cHigh$:
\begin{eqnarray}
\label{Eq:theta-u+}
\hat \theta(u_+) \neq 0 \Rightarrow\quad &
0
<m_\cHigh<
m_{cb}^{\max}, &  \label{eq:thup} \\[3mm] 
\hat \theta(v_+) \neq 0 \Rightarrow \quad & \displaystyle
\left[\frac{a (m_{ca}^{\max})^{2}}{(m_{cb}^{\max})^2}+1 \right] m_\cHigh^2
-(m_{ca}^{\max})^{2} 
<x<
m_\cHigh^2, & \label{eq:thvp} \\[3mm]
\hat \theta(u_-) \neq 0 \Rightarrow \quad &
-m_\cHigh^{2}
<x<
(m_{cb}^{\max})^{2}-m_\cHigh^{2}, & \label{eq:thum} \\[3mm]
\hat \theta(v_-) \neq 0 \Rightarrow\quad & \displaystyle
x< \frac{(m_{cb}^{\max})^2}{a} 
- \left[\frac{(m_{cb}^{\max})^2}{a (m_{ca}^{\max})^2}+1 \right] m_\cHigh^2.
& \label{eq:thvm}
\end{eqnarray}
The first two inequalities constrain the integration over positive values of
$x$, while the last two constrain the integration over negative values of
$x$. Notice that $v_-$ only provides one inequality since the condition
$\theta(v_-)\neq 0$ holds when (\ref{eq:thum}) holds.  Also $\theta(v_+)\neq
0$ yields an upper bound on $x$ because the denominator of $v_+$ is always
positive when (\ref{Eq:theta-u+}) holds.

\ssp

Which of these inequalities provides the strongest bound and thereby
the endpoint of the integration is highly dependent on the mass
hierarchy between particles $A$, $B$ and $C$. In particular, the
various bounds on $x$ coincide at four (non-trivial) distinct values
of $m_\cHigh$, namely $m_{ca}^{\max}$, $m_{cb}^{\max}$,
$\frac{m_B}{m_C} m_{ca}^{\max}$ and $m_{c2{\rm (eq)}}^{\max}$, and it
is the relative size of these four quantities which is important.
With this in mind we define three different regions, exhausting all
possible hierarchies.  Writing
\begin{equation}
R_A \equiv \frac{m_A^2}{m_B^2}, \quad 
R_B \equiv \frac{m_B^2}{m_C^2}, \quad 
R_C \equiv \frac{m_C^2}{m_D^2},
\label{eq:ratios-R}
\end{equation}
($R_C$ is defined for later reference), the different mass hierarchies
can be divided into:
\begin{equation}
\begin{array}{lrcccl}
\text{Region~1:} \qquad & \frac{1}{2-R_A} &<& R_B &<& 1, \nonumber \\
\text{Region~2:} & R_A &<& R_B &<& \frac{1}{2-R_A}, \nonumber \\
\text{Region~3:} & 0 &<& R_B &<& R_A.
\end{array}
\label{eq:cases}
\end{equation}
The three regions are shown in figure~\ref{fig:limits} over the space of
$R_A$ and $R_B$.

\FIGURE[htb]{
\epsfxsize 9cm
\mbox{\epsfbox{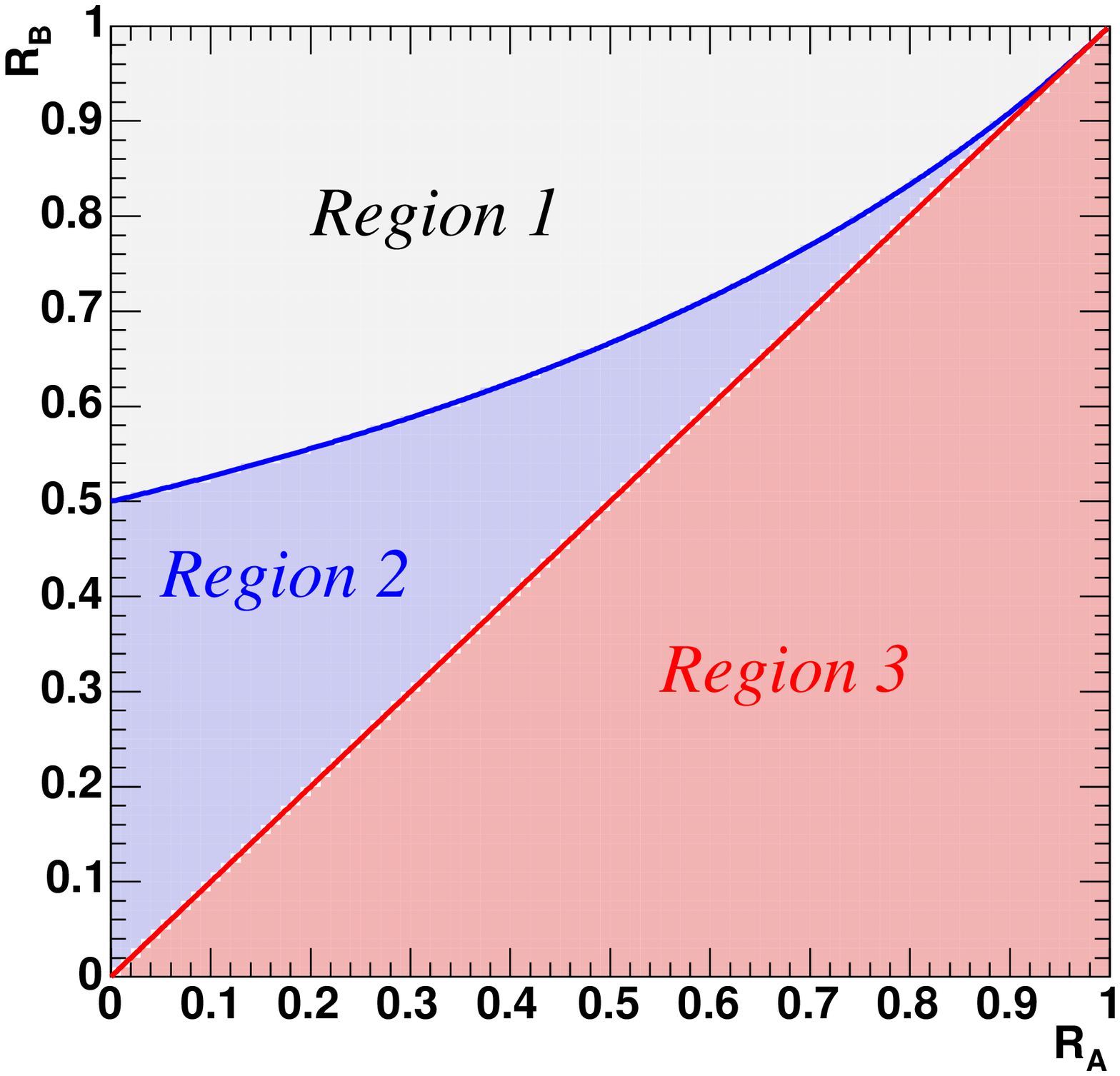}}
\caption{Regions 1, 2, and 3 vs.\ $R_A \equiv (m_A/m_B)^2$ and $R_B \equiv
(m_B/m_C)^2$.
\label{fig:limits}}
}

\subsubsection{Region 1: $\frac{1}{2-R_A} < R_B < 1$}

In Region 1, we have
\begin{equation}
m_{cb}^{\max} < m_{c2{\rm (eq)}}^{\max} < \frac{m_B}{m_C} m_{ca}^{\max} 
< m_{ca}^{\max}.
\label{eq:region1_ineq}
\end{equation}
The integration over positive values of $x$ is restricted by
eq.~(\ref{eq:thvp}), and is non-zero only when $m_\cHigh$ satisfies
eqs.~(\ref{eq:thup}).  If the lower bound on $x$ is greater than zero,
it will provide the lower limit of the integration. However, this
is only the case when
\begin{equation}
m_\cHigh  > m_{c2{\rm (eq)}}^{\max}
\label{eq:mchigh.gt.meq}
\end{equation}
which can never be true for Region~1 due to eq.~(\ref{eq:thup}), and there is
only one permitted range:
\begin{equation}
0 < x < m_\cHigh^2 \qquad \text{for} \qquad 0<m_\cHigh<m_{cb}^{\max}.
\end{equation}

\ssp

For the integral over negative values, $x$ is restricted by
eqs.~(\ref{eq:thum}) \& (\ref{eq:thvm}). While the lower bound is
unambiguous there are three possible upper bounds (including the
original $x<0$ from the integral) and we must divide the solution into
three possible cases depending on which upper bound is dominant:
\begin{equation}
\begin{array}{lclcl}
-m_\cHigh^{2} & < x < & 0 
& \text{for~~} & 
0<m_\cHigh<m_{cb}^{\max}, \\[3mm]
-m_\cHigh^{2} & < x < & (m_{cb}^{\max})^{2}-m_\cHigh^{2} \hspace{26mm}
& \text{for~~} & 
m_{cb}^{\max}<m_\cHigh
<\frac{m_{B}}{m_{C}}m_{ca}^{\max}, \\[3mm]
\lefteqn{-m_\cHigh^{2} < x < \frac{(m_{cb}^{\max})^{2}}{a}
-\left[ \frac{(m_{cb}^{\max})^{2}}{a(m_{ca}^{\max})^{2}}
+1 \right] m_\cHigh^{2}} &&&& \\[2mm]
&&& \text{for~~} & 
\frac{m_{B}}{m_{C}}
m_{ca}^{\max}<m_\cHigh<m_{ca}^{\max}.
\end{array}
\label{eq:region1_xneg}
\end{equation}

\ssp

Performing the two integrals in the appropriate cases and adding them
together we find the full expression for the $m_\cHigh^2$
distribution:
\begin{equation}
\frac{1}{\Gamma_0} \frac{\partial \Gamma_0}{\partial m_\cHigh^2} = \left\{
\begin{array}{lcl} \displaystyle
\lefteqn{\frac{1}{(m_{ca}^{\max})^{2}a}
\left[
\ln\frac{(m_{cb}^{\max})^{2}}
{(m_{cb}^{\max})^{2}-am_\cHigh^{2}}
+\frac{am_\cHigh^{2}}{(m_{cb}^{\max})^{2}-am_\cHigh^{2}}
\right]} && \\[6mm]
&\text{for~~}& 0<m_\cHigh<m_{cb}^{\max}, \\[5mm] \displaystyle
\frac{1}{(m_{ca}^{\max})^{2}a}
\ln\frac{m_{C}^{2}}{m_{B}^{2}} &\text{for~~}&
m_{cb}^{\max}<m_\cHigh<\frac{m_{B}}{m_{C}}
m_{ca}^{\max}, \\[5mm] \displaystyle
\frac{1}{(m_{ca}^{\max})^{2}a}
\ln\frac{(m_{ca}^{\max})^{2}}{m_\cHigh^{2}} \hspace{1.6cm} &\text{for~~} &
\frac{m_{B}}{m_{C}}m_{ca}^{\max}<m_\cHigh<m_{ca}^{\max},
\end{array} \right. 
\end{equation}
and zero otherwise.

\subsubsection{Region~2: $R_A < R_B < \frac{1}{2-R_A}$}

For Region 2 we have
\begin{equation}
\frac{m_B}{m_C} m_{ca}^{\max} 
<
m_{c2{\rm (eq)}}^{\max} 
<
m_{cb}^{\max} 
< 
m_{ca}^{\max}.
\end{equation}
As for Region~1, the integration over positive $x$ is restricted
by eq.~(\ref{eq:thvp}). However, {\it unlike} Region~1, the lower limit
of eq.~(\ref{eq:thvp}) can now be larger than zero, providing a new
lower limit of integration. This happens when
eq.~(\ref{eq:mchigh.gt.meq}) is satisfied, and provides different
limits of integration for two distinct cases:
\begin{equation}
\begin{array}{rclcl}
0 & < x < & m_\cHigh^2 
& \text{~~for~~} & 
0<m_\cHigh<m_{c2{\rm (eq)}}^{\max}, \\[2mm]
\displaystyle
\left[\frac{a (m_{ca}^{\max})^{2}}{(m_{cb}^{\max})^2}+1 \right] m_\cHigh^2
-(m_{ca}^{\max})^{2}
& < x < & m_\cHigh^2 
& \text{~~for~~} & 
m_{c2{\rm (eq)}}^{\max}<m_\cHigh
<m_{cb}^{\max}.
\end{array}
\label{eq:region2_limits_pos}
\end{equation}

\ssp

For the integration over negative $x$, the step-functions again
require that eq.~(\ref{eq:thum}) and eq.~(\ref{eq:thvm}) hold, if the
integral is to be non-zero. However, since $m_{cb}^{\max}$ is larger
than $\frac{m_B}{m_C} m_{ca}^{\max}$ in Region~2, the upper bound from
eq.~(\ref{eq:thum}) is never dominant. Therefore Region~2 gives
only two different sets of integration limits:
\begin{equation}
\begin{array}{rclcl}
-m_\cHigh^2 & < x < & 0
& \hspace{5cm} \text{~~for~~} & 
0<m_\cHigh< m_{c2{\rm (eq)}}^{\max},\\[2mm] 
-m_\cHigh^2 & < x < & \lefteqn{\displaystyle 
\frac{(m_{cb}^{\max})^2}{a} 
- \left[\frac{(m_{cb}^{\max})^2}{a (m_{ca}^{\max})^2}+1 \right] m_\cHigh^2} 
&& \\[2mm] 
&&&  \hspace{5cm} \text{~~for~~} & 
m_{c2{\rm (eq)}}^{\max}<m_\cHigh
<m_{ca}^{\max}.
\end{array}
\label{eq:region2_limits_neg}
\end{equation}

\ssp

For Region~2 the full distribution is then given by
\begin{equation}
\frac{1}{\Gamma_0} \frac{\partial \Gamma_0}{\partial m_\cHigh^2} = \left\{
\begin{array}{lcl} \displaystyle
\lefteqn{\frac{1}{(m_{ca}^{\max})^{2}a}
\left[
\ln\frac{(m_{cb}^{\max})^{2}}{(m_{cb}^{\max})^{2}-am_{\cHigh}^{2}}
+\frac{am_{\cHigh}^{2}}{(m_{cb}^{\max})^{2}-am_\cHigh^{2}}
\right]} \hspace{60mm}  &&\\[5mm]
&\text{for~~}& 
0<m_{\cHigh}<m_{c2{\rm (eq)}}^{\max},
\\[5mm] \displaystyle
\frac{1}{(m_{ca}^{\max})^{2}a}
\left[
\ln\frac{(m_{ca}^{\max})^{2}}{m_{\cHigh}^{2}}
+\frac{a(m_{ca}^{\max})^{2}}{(m_{cb}^{\max})^{2}}
\right]
&\text{for~~}&
m_{c2{\rm (eq)}}^{\max}<m_{\cHigh}
<m_{cb}^{\max},
\\[5mm] \displaystyle
\frac{1}{(m_{ca}^{\max})^{2}a}
\ln\frac{(m_{ca}^{\max})^{2}}{m_{\cHigh}^{2}}
&\text{for~~} &
m_{cb}^{\max}<m_{\cHigh}<m_{ca}^{\max},
\end{array} \right. 
\end{equation}
and zero otherwise.

\subsubsection{Region~3: $0 < R_B < R_A$}

In Region~3 we have
\begin{equation}
\frac{m_{B}}{m_{C}}m_{ca}^{\max}
<m_{c2{\rm (eq)}}^{\max}
<m_{ca}^{\max}<m_{cb}^{\max}.
\end{equation}
For the individual integrations over positive and negative $x$,
Region~3 is identical to Region~2, and we again obtain
eqs.~(\ref{eq:region2_limits_pos}) \& (\ref{eq:region2_limits_neg}) for
the positive and negative integration limits respectively. However,
since $m_{cb}^{\max}$ is now {\it larger} then $m_{ca}^{\max}$,
contrary to Region~2, the sum of the two contributions will be
different. We find:
\begin{equation}
\frac{1}{\Gamma_0} \frac{\partial \Gamma_0}{\partial m_\cHigh^2} = \left\{
\begin{array}{lcl} \displaystyle
\lefteqn{\frac{1}{(m_{ca}^{\max})^{2}a}
\left[
\ln\frac{(m_{cb}^{\max})^{2}}{(m_{cb}^{\max})^{2}-am_{\cHigh}^{2}}
+\frac{am_{\cHigh}^{2}}{(m_{cb}^{\max})^{2}-am_\cHigh^{2}}
\right]} \hspace{60mm}  &&\\[5mm]
&\text{for~~}& 0<m_{\cHigh}<m_{c2{\rm (eq)}}^{\max},
\\[5mm] \displaystyle
\frac{1}{(m_{ca}^{\max})^{2}a}
\left[
\ln\frac{(m_{ca}^{\max})^{2}}{m_{\cHigh}^{2}}
+\frac{a(m_{ca}^{\max})^{2}}{(m_{cb}^{\max})^{2}}
\right]
&\text{for~~}& m_{c2{\rm (eq)}}^{\max}<m_{\cHigh}<m_{ca}^{\max},
\\[5mm] \displaystyle
\frac{1}{(m_{cb}^{\max})^{2}}
&\text{for~~} &
m_{ca}^{\max}<m_{\cHigh}<m_{cb}^{\max},
\end{array} \right. 
\end{equation}
and zero otherwise.

\subsection{The two-particle invariant mass $m_{\cLow}$}
\label{sect:cLow}

The invariant mass $m_{\cLow}$ is given by
\begin{equation}
m_{\cLow}=\min\left(m_{cb},m_{ca}\right).
\end{equation}
For calculation of the differential decay rate we can adopt a
method very similar to that of $m_\cHigh$ above. Using $u$
and $v$ as defined in eq.~(\ref{eq:indep-var}) we can write $m_{\cLow}$
as
\begin{equation}
m_{\cLow}^2
=\min\left[(m_{cb}^{\max})^{2}u,
(m_{ca}^{\max})^{2}\left(1-au\right)v\right].
\label{eq:mc2Low}
\end{equation}
Keeping $x$ as defined in eq.~(\ref{eq:def-x}), we write
\begin{equation}
m_\cLow^{2}=\theta(-x)(m_{cb}^{\max})^{2}u
+\theta(x)(m_{ca}^{\max})^{2}\left(1-au\right)v.
\end{equation}
These can be inverted to give
\begin{equation}
u=\frac{m_\cLow^{2}+\theta(x)x}{(m_{cb}^{\max})^{2}},\qquad
v=\frac{m_\cLow^{2}-\theta\left(-x\right)x}
{(m_{ca}^{\max})^{2}\left(1-a\frac{m_\cLow^{2}+\theta(x)x}
{(m_{cb}^{\max})^{2}}\right)}, 
\end{equation}
and the differential distribution again has the form 
(\ref{Eq:dist-c2-high}), but now with
\begin{eqnarray}
\lefteqn{u_- = \frac{m_\cLow^{2}}{(m_{cb}^{\max})^{2}},} \hspace{6.5cm} &&
u_+ = \frac{m_\cLow^{2}+x}{(m_{cb}^{\max})^{2}}, \\
\lefteqn{v_- = \frac{m_\cLow^{2}-x}
{(m_{ca}^{\max})^{2}\left(1-a\frac{m_\cLow^{2}}
{(m_{cb}^{\max})^{2}}\right)},} \hspace{6.5cm} &&
v_+ = \frac{m_\cLow^{2}}
{(m_{ca}^{\max})^{2} \left(1-a\frac{m_\cLow^{2}+x}
{(m_{cb}^{\max})^{2}}\right)}.
\end{eqnarray}
These are actually the same definitions as eqs.~(\ref{eq:upm}) and
(\ref{eq:vpm}) only now written in terms of $m_\cLow$ rather than
$m_\cHigh$, using
\begin{equation}
|x|=m_\cHigh^{2}-m_\cLow^{2},
\end{equation}
as follows from (\ref{eq:mc2High}), (\ref{eq:def-x}) and
(\ref{eq:mc2Low}). The step-functions of the integrand then give
the following restrictions on $x$ and/or $m_\cLow$:
\begin{eqnarray}
\hat \theta(u_+) \neq 0 \Rightarrow\quad &
-m_\cLow^{2}
<x<
(m_{cb}^{\max})^{2}-m_\cLow^{2}, & \label{eq:thup2} \\[3mm] 
\hat \theta(v_+) \neq 0 \Rightarrow \quad & \displaystyle
x< \frac{(m_{cb}^{\max})^2}{a} 
- \left[\frac{(m_{cb}^{\max})^2}{a (m_{ca}^{\max})^2}+1 \right] m_\cLow^2,
& \label{eq:thvp2} \\[3mm]
\hat \theta(u_-) \neq 0 \Rightarrow \quad &
0<m_\cLow <m_{cb}^{\max},
& \label{eq:thum2}\\[3mm]
\hat \theta(v_-) \neq 0 \Rightarrow\quad & \displaystyle
\left[\frac{a (m_{ca}^{\max})^{2}}{(m_{cb}^{\max})^2}+1 \right] m_\cLow^2
-(m_{ca}^{\max})^{2} 
<x<
m_\cLow^2. & \label{eq:thvm2} 
\end{eqnarray}
The parallels with the $m_\cHigh$ case are obvious. The step-functions
provide the same constraints for $m_\cLow$ as they did for $m_\cHigh$,
but constraints which were previously for negative $x$ are now for
positive $x$ and vice versa. Therefore we must be careful with the
$x=0$ boundary: we will have the same regions of applicability as
shown in figure~\ref{fig:limits} but the inequalities giving the
integration limits for each region will be different from the
$m_\cHigh$ case.

\subsubsection{Region 1: $\frac{1}{2-R_A} < R_B < 1$}

In Region~1, eq.~(\ref{eq:region1_ineq}) holds as before. 

\ssp

For positive $x$, we have constraints given by eqs.~(\ref{eq:thup2}) \&
(\ref{eq:thvp2}). However, in Region~1 the upper bound on $x$ provided
by eq.~(\ref{eq:thup2}) is always more restrictive, and insists that
$m_{\cLow}$ be smaller than $m_{cb}^{\max}$ for a nonzero
result. Finally, since $m_\cLow^2$ is necessarily positive, the lower
bound on $x$ is trivially zero. Therefore, the integration limits for
Region~1 with positive $x$ only has one case:
\begin{equation}
0 < x < (m_{cb}^{\max})^2 - m_{\cLow}^2 
\text{~~for~~}  
0 < m_{\cLow} < m_{cb}^{\max}.
\end{equation}

\ssp

For negative $x$, the integration limits also only have one case since
eq.~(\ref{eq:thum2}) restricts $m_{\cLow}$ to be below all the
characteristic masses in  eq.~(\ref{eq:region1_ineq}). This gives
\begin{equation}
-(m_{ca}^{\max})^{2} 
+ \left[\frac{a (m_{ca}^{\max})^{2}}{(m_{cb}^{\max})^2}+1 \right] m_\cLow^2
<x<0
\text{~~for~~}  
0 < m_{\cLow} < m_{cb}^{\max}.
\end{equation}

\ssp

It is then rather trivial to calculate the full integral for Region~1:
\begin{equation}
\frac{1}{\Gamma_0} \frac{\partial \Gamma_0}{\partial m_\cLow^2} =
\frac{1}{(m_{ca}^{\max})^{2}a}
\left[
\ln\frac{(m_{cb}^{\max})^{2}-am_\cLow^{2}}
{\frac{m_B^2}{m_C^2}(m_{cb}^{\max})^{2}}
+\frac{a(m_{ca}^{\max})^{2}}{(m_{cb}^{\max})^{2}}
-\frac{am_\cLow^{2}}{(m_{cb}^{\max})^{2}-am_\cLow^{2}}
\right],
\end{equation}
for $0<m_\cLow<m_{cb}^{\max}$, and zero otherwise.

\subsubsection{Regions 2 and 3: $0 < R_B < \frac{1}{2-R_A}$}

The inequalities given by the step-functions together with the
positivity or negativity constraints on $x$, are independent of the
boundary between Regions 2 and 3. In other words, which of
$m_{ca}^{\max}$ and $m_{cb}^{\max}$ is larger is irrelevant when both
are larger than $m_{c2{\rm (eq)}}^{\max}$. Therefore the analytic form
of the distribution will be the same in Region 2 as it is in Region 3
and we do not need to treat them separately. The hierarchy
of characteristic masses is then:
\begin{equation}
\frac{m_B}{m_C} m_{ca}^{\max} < m_{c2{\rm (eq)}}^{\max} 
< \min(m_{ca}^{\max},m_{cb}^{\max})
\end{equation}

For positive $x$, the issue is which of the upper bounds of
eqs.~(\ref{eq:thup2}) \& (\ref{eq:thvp2}) is more restrictive. For
higher values of $m_\cLow$ it is eq.~(\ref{eq:thvp2}) which is more
restrictive, with the transition being at $\frac{m_B}{m_C}
m_{ca}^{\max}$ as already intimated by eq.~(\ref{eq:region1_xneg}),
and we must also ensure that this upper bound on $x$ is
positive. Together, these considerations give the integration limits:
\begin{equation}
\begin{array}{rclcl}
0 & < x < & (m_{cb}^{\max})^2-m_\cLow^2
& \text{~~for~~} & 
0<m_\cLow< \frac{m_B}{m_C} m_{ca}^{\max},
\\[2mm] 
0 & < x < & \frac{(m_{cb}^{\max})^2}{a} 
- \left[\frac{(m_{cb}^{\max})^2}{a (m_{ca}^{\max})^2}+1 \right] m_\cLow^2 
& \text{~~for~~} & 
\frac{m_B}{m_C} m_{ca}^{\max} < m_\cLow < m_{c2{\rm (eq)}}^{\max}.
\end{array}
\end{equation}

\ssp

For negative $x$, only the (lower) constraint of eq.~(\ref{eq:thvm2})
is interesting. The insistence that this lower bound is negative
requires that $m_\cLow$ is smaller than $m_{c2{\rm (eq)}}^{\max}$,
which is more restrictive than eq.~(\ref{eq:thum2}) for these regions. 
The integration limits for negative $x$ are therefore again rather simple:
\begin{equation}
-(m_{ca}^{\max})^{2} 
+ \left[\frac{a (m_{ca}^{\max})^{2}}{(m_{cb}^{\max})^2}+1 \right] m_\cLow^2
<x<0 
\text{~~for~~} 
0 < m_\cLow < m_{c2{\rm (eq)}}^{\max}.
\end{equation}

\ssp

The full result for Regions 2 and 3 is:
\begin{align}
&\frac{1}{\Gamma_0} \frac{\partial \Gamma_0}{\partial m_\cLow^2} = 
\nonumber \\[5mm]
&\left\{
\begin{array}{lcl} \displaystyle
\lefteqn{\frac{1}{(m_{ca}^{\max})^{2}a}
\left[
\ln\frac{(m_{cb}^{\max})^2-am_\cLow^2}{\frac{m_B^2}{m_C^2}(m_{cb}^{\max})^2}
+\frac{a(m_{ca}^{\max})^{2}}{(m_{cb}^{\max})^{2}}
-\frac{am_\cLow^2}{(m_{cb}^{\max})^{2}-am_\cLow^2}
\right]}
 \hspace{61mm}  &&\\[5mm]
&\text{for~~}& 
0<m_{\cLow}<\frac{m_{B}}{m_{C}}m_{ca}^{\max},
\\[5mm] \displaystyle
\lefteqn{\frac{1}{(m_{ca}^{\max})^{2}a}
\left[
\ln\frac{(m_{ca}^{\max})^{2}\left[(m_{cb}^{\max})^{2}-am_\cLow^2\right]}
{(m_{cb}^{\max})^{2}m_\cLow^2}
+\frac{a(m_{ca}^{\max})^{2}}{(m_{cb}^{\max})^{2}}
-\frac{am_\cLow^2}{(m_{cb}^{\max})^{2}-am_\cLow^2}
\right]} \\[5mm]
&\text{for~~}&
\frac{m_{B}}{m_{C}}m_{ca}^{\max}< m_\cLow <
m_{c2{\rm (eq)}}^{\max}, 
\end{array} \right. \nonumber
\\[2mm]
\end{align}
and zero otherwise.

\subsection{The three-particle invariant mass $m_{cba}$}

The quantity $m_{cba}$ is defined by
\begin{equation}
m_{cba}^2 \equiv (p_c+p_b+p_a)^2.
\end{equation}
Making use of energy and momentum conservation, we can write this in
terms of particle masses and two angles:
\begin{eqnarray}
m_{cba}^2 &=& 
m_{D}^{2}-m_{B}^{2}-y
\left(\frac{m_{B}^{2}+m_{A}^{2}}{m_{B}}\right) \nonumber \\
&&+\sqrt{y^2+ 2 y m_{B}+m_{B}^{2}-m_{D}^{2}}
\left(\frac{m_{B}^{2}-m_{A}^{2}}{m_{B}}\right)\left(1-2w\right),
\end{eqnarray}
where $u$ is given by eq.~(\ref{eq:indep-var}), $w$ is similarly
defined in terms of the angle between $A$ and $D$ in the rest frame of
$B$,
\begin{equation}
w \equiv \frac{1-\cos\theta_{AD}^{(B)}}{2},
\end{equation}
and the quantity $y$ is defined by
\begin{equation}
y \equiv \frac{\left(m_{D}^{2}-m_{B}^{2}\right)m_{B}}{2m_{C}^{2}}u
+\frac{\left(m_{C}^{2}-m_{B}^{2}\right)^{2}}{2m_{B}m_{C}^{2}}u
+\left(\frac{m_{D}^{2}-m_{B}^{2}}{2m_{B}}\right)(1-u).
\end{equation}
It is straightforward to solve for $u$ and $w$, 
\begin{align}
u&=\frac{\left(m_{D}^{2}-m_{B}^{2}\right)m_{C}^{2}-2m_{C}^{2}m_{B}y}
{\left(m_{D}^{2}-m_{C}^{2}\right)\left(m_{C}^{2}-m_{B}^{2}\right)},
\nonumber \\[4mm]
w&=\frac{m_{D}^{2}-m_{B}^{2}
-y\left(\frac{m_{B}^{2}+m_{A}^{2}}{m_{B}}\right)
+\sqrt{y^{2}+2ym_{B}+m_{B}^{2}-m_{D}^{2}}
\left(\frac{m_{B}^{2}-m_{A}^{2}}{m_{B}}\right)-m_{cba}^{2}}
{2\sqrt{y^{2}+2ym_{B}+m_{B}^{2}-m_{D}^{2}}
\left(\frac{m_{B}^{2}-m_{A}^{2}}{m_{B}}\right)}.
\end{align} 

\ssp

Under the assumption that $B$ is a scalar particle, its decay is
istotropic and the doubly-differential width in terms of these
two variables $u$ and $w$ must be flat. Therefore we can write
\begin{eqnarray}
\frac{1}{\Gamma_0} \frac{\partial^2 \Gamma_0}{\partial y \partial m_{cba}^2}
&=&\left| \frac{\partial (u,w)}{\partial (y,m_{cba}^2)} \right|
\frac{1}{\Gamma_0} \frac{\partial^2 \Gamma_0}{\partial u \partial w}
\nonumber \\
&=& 
\frac{m_{C}^{2}m_{B}^{2}}
{\left(m_{D}^{2}-m_{C}^{2}\right)\left(m_{C}^{2}-m_{B}^{2}\right)
\left(m_{B}^{2}-m_{A}^{2}\right)}
\frac{\hat \theta(u) \hat \theta(w)}{\sqrt{y^{2}+2ym_{B}+m_{B}^{2}-m_{D}^{2}}},
\end{eqnarray}
and integrate over $y$ to acquire the desired distribution,
\begin{eqnarray}
\frac{1}{\Gamma_0} \frac{\partial \Gamma_0}{\partial m_{cba}^2}
&=&
\frac{1}
{\left(m_{ca}^{\text{max}}\right)^2a} \int_{-\infty}^{\infty} 
\frac{\hat \theta(u) \hat \theta(w)}
{\sqrt{y^{2}+2ym_{B}+m_{B}^{2}-m_{D}^{2}}} dy .
\nonumber \\
\end{eqnarray}

\ssp

At this point it is useful to make some definitions, in order to help
keep expressions compact. Let us first define the integral:
\begin{eqnarray}
L(a_1,a_2) & \equiv & 
\int_{a_{1}}^{a_{2}}
\frac{1}{\sqrt{y^{2}+2ym_{B}+m_{B}^{2}-m_{D}^{2}}}dy \nonumber \\
&=&\ln\frac{a_{2}+m_{B}+\sqrt{a_{2}^{2}+2a_{2}m_{B}+m_{B}^{2}-m_{D}^{2}}}
{a_{1}+m_{B}+\sqrt{a_{1}^{2}+2a_{1}m_{B}+m_{B}^{2}-m_{D}^{2}}},
\label{eq:define-L}
\end{eqnarray}
which is the only integral that we will need.  The mass values which
appear as kinematical endpoints of $m_{cba}$ are
\cite{Allanach:2000kt,Gjelsten:2004}:
\begin{align}
m_{1}^{2} &=\frac{\left(m_{D}^{2}-m_{C}^{2}\right)
\left(m_{C}^{2}-m_{A}^{2}\right)}{m_{C}^{2}},
\label{eq:m_1^2}  \\
m_{2}^{2}
&=\frac{\left(m_{C}^{2}-m_{B}^{2}\right)
\left(m_{D}^{2}m_{B}^{2}-m_{C}^{2}m_{A}^{2}\right)}
{m_{B}^{2}m_{C}^{2}}, 
\label{eq:m_2^2} \\
m_{3}^{2}
&=\frac{\left(m_{D}^{2}-m_{B}^{2}\right)
\left(m_{B}^{2}-m_{A}^{2}\right)}{m_{B}^{2}}, 
\label{eq:m_3^2} \\
m_{4}^{2}&=\left(m_{D}-m_{A}\right)^{2}.
\label{eq:m_4^2}
\end{align}
The mass $m_{4}$ is always the largest, as is seen by 
\begin{eqnarray}
m_{4}^{2}-m_{1}^{2} &=&
\left(m_{C}^{2}-m_{A}m_{D}\right)^{2} / m_{C}^{2}, \nonumber \\
m_{4}^{2}-m_{2}^{2} &=&
\left(m_{B}^{2}m_{D}-m_{A}m_{C}^{2}\right)^{2} / (m_{B}^{2}m_{C}^{2}), 
\nonumber \\
m_{4}^{2}-m_{3}^{2} &=&
\left(m_{B}^{2}-m_{A}m_{D}\right)^{2} / m_{B}^{2} . 
\label{eq:m_4-order}
\end{eqnarray}
For the others, the relative order will depend on the values
of $m_{A}$, $m_{B}$, $m_{C}$, and $m_{D}$:
\begin{eqnarray}
m_{1}^{2}-m_{2}^{2}
&=&\left(m_{B}^{2}-m_{A}^{2}\right)
\left(m_{B}^{2}m_{D}^{2}-m_{C}^{2}m_{C}^{2}\right)/(m_{B}^{2}m_{C}^{2}), 
\nonumber \\
m_{1}^{2}-m_{3}^{2}
&=&\left(m_{C}^{2}-m_{B}^{2}\right)
\left(m_{A}^{2}m_{D}^{2}-m_{B}^{2}m_{C}^{2}\right)/(m_{B}^{2}m_{C}^{2}), 
\nonumber \\
m_{2}^{2}-m_{3}^{2}
&=&\left(m_{D}^{2}-m_{C}^{2}\right)
\left(m_{A}^{2}m_{C}^{2}-m_{B}^{2}m_{B}^{2}\right)/(m_{B}^{2}m_{C}^{2}).
\label{eq:m_order}
\end{eqnarray}

\ssp

As for the previous distributions, it is important to determine the
upper and lower integration bounds. The step-functions for $u$ provide
rather simple constraints,
\begin{equation}
y_{1}
\equiv \frac{m_{D}^{2}-m_{B}^{2}
-(m_{cb}^{\max})^{2}}{2m_{B}}
< y < \frac{m_{D}^{2}-m_{B}^{2}}{2m_{B}} \equiv y_{2},
\label{eq:y12}
\end{equation}
where we have defined the quantities $y_{1,2}$ for later
convenience. The constraints from the step-functions for $w$ are
somewhat more complicted,
\begin{equation}
y_{3} < y < y_{4},
\end{equation}
where
\begin{eqnarray}
y_{4,3}&\equiv& \frac{1}{4m_{B}m_{A}^{2}} \left\{
\lefteqn{\phantom{\sqrt{\left(m_{D}^{2}-m_{A}^{2}-m_{cba}^{2}\right)^{2}}}}
\left(m_{B}^{2}-m_{A}^{2}\right)^{2}
-\left[m_{cba}^{2}-\left(m_{D}^{2}-m_{B}^{2}\right)\right]
\left(m_{B}^{2}+m_{A}^{2}\right)  \right. \nonumber \\ 
&& \left. \qquad \qquad \pm \left(m_{B}^{2}-m_{A}^{2}\right)
\sqrt{\left(m_{D}^{2}-m_{A}^{2}-m_{cba}^{2}\right)^{2}
-4m_{A}^{2}m_{cba}^{2}} \,
\right\} .
\label{eq:y34}
\end{eqnarray}
In order that $y_{4,3}$ be real, we must insist that
$m_{cba} < m_{4}$, with $m_{4}$ defined by eq.~(\ref{eq:m_4^2}).

We now need to compare these constraints and determine the relative
ordering of $y_1$, $y_2$, $y_3$ and $y_4$, to see which will form the
endpoints of the $y$ integration. There are four possible cases where
the constraints overlap, as illustrated in figure~\ref{fig:ranges}.

\ssp

\noindent {\bf \underline{$y_1$ vs.\ $y_4$}:} In order to have a
physical solution, we require that $y_1<y_4$. This constraint is
manifest differently for different mass hierarchies. Comparing the
expressions from eqs.~(\ref{eq:y12}) \& (\ref{eq:y34}) we find the
requirements,
\begin{equation}
\begin{array}{lcrcl}
{\rm for~~} m_C^2<m_Am_D \:& (\Rightarrow& m_Am_C^2<m_B^2m_D), & \: & 
 y_1<y_4 {\rm ~if~} m_{cba}<m_1, \\
{\rm for~~} m_C^2>m_Am_D \:&{\rm \&}& m_Am_C^2>m_B^2m_D, \phantom{)} &  \:& 
 y_1<y_4 {\rm ~if~} m_{cba}<m_2, \\
{\rm for~~} m_C^2>m_Am_D \:&{\rm \&}& m_Am_C^2<m_B^2m_D, \phantom{)} &  \:& 
 y_1<y_4 {\rm ~if~} m_{cba}<m_4,
\end{array}
\end{equation}
where $m_1$, $m_2$ and $m_4$ are given by eqs.~(\ref{eq:m_1^2}),
(\ref{eq:m_2^2}) \& (\ref{eq:m_4^2}) respectively.

\FIGURE[ht]{
\epsfxsize 6.5cm
\epsfbox{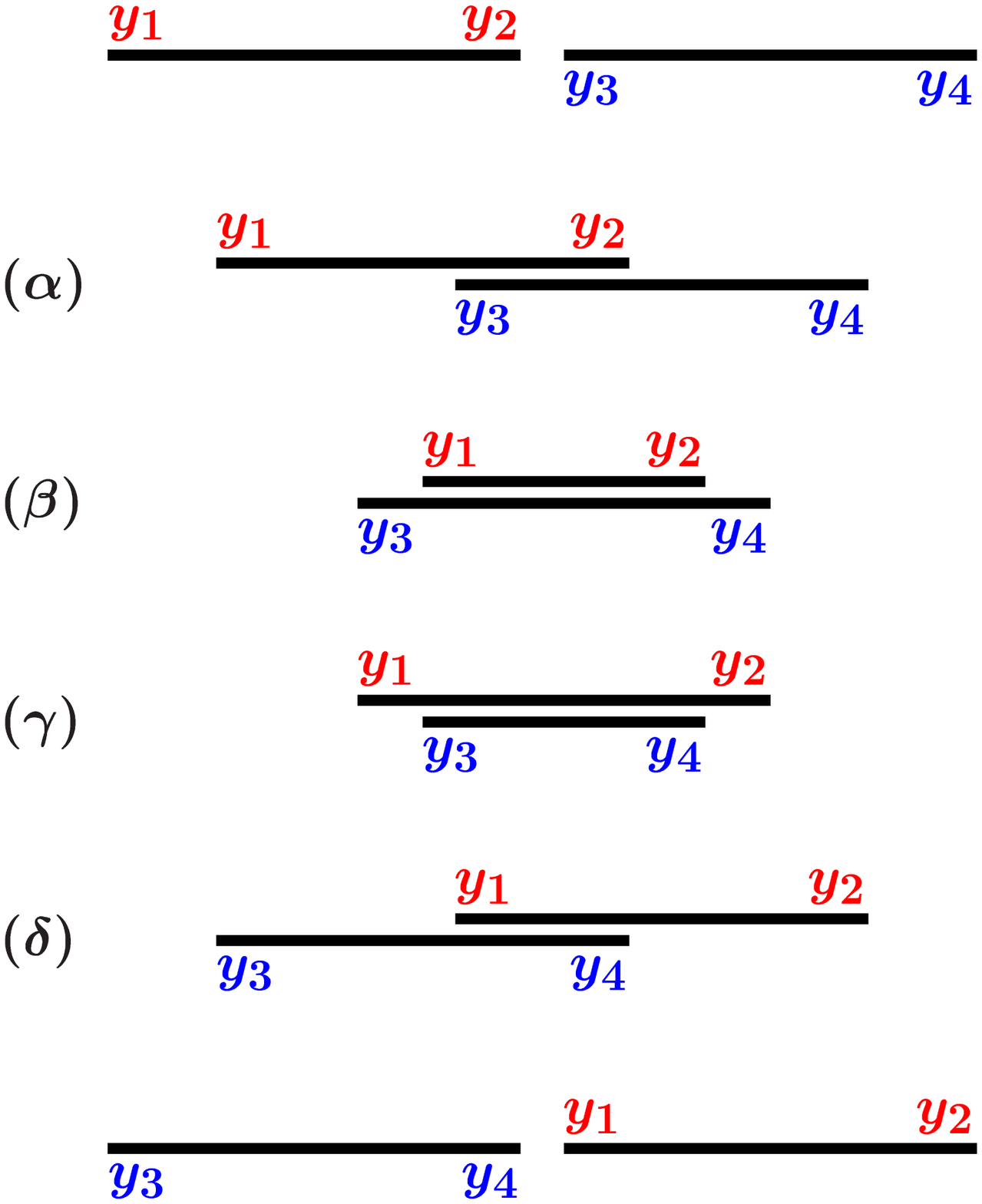}
\caption{Schematic representation of integration ranges.  The cases
($\alpha$)--($\delta$) give non-zero contributions.}
\label{fig:ranges}
}

\ssp

\noindent {\bf \underline{$y_2$ vs.\ $y_3$}:} Similarly, to provide an overlap
of the integration regions we must insist that $y_2>y_3$. Once again,
which values of $m_{cba}$ are required is dependent on the mass
hierarchy:
\begin{equation}
\begin{array}{lclcl}
{\rm for~~} m_B^2<m_Am_D, & \: & 
 y_3<y_2 {\rm ~if~} m_{cba}<m_4, \\
{\rm for~~} m_B^2>m_Am_D, & \: & 
 y_3<y_2 {\rm ~if~} m_{cba}<m_3,
\end{array}
\end{equation}
where $m_3$ is given by eq.~(\ref{eq:m_3^2}).

\ssp

\noindent {\bf \underline{$y_2$ vs.\ $y_4$}:} Although one does not require any
particular relation between $y_2$ and $y_4$ in order to have a
non-zero result, only the smallest of $y_2$ and $y_4$ will provide the
upper bound on the integration, so it is important to verify which
mass regimes lead to which dominant upper bound. We find:
\begin{equation}
\begin{array}{lcrcl}
{\rm for~~} m_B^2<m_Am_D, & \: & 
 y_2<y_4 {\rm ~if~} m_{cba}<m_3, \\
{\rm for~~} m_B^2>m_Am_D, & \: & 
 y_2<y_4 {\rm ~if~} m_{cba}<m_4.
\end{array}
\end{equation}

\ssp

\noindent {\bf \underline{$y_1$ vs.\ $y_3$}:} Finally, the lower bound of the
integration is governed by whichever of $y_1$ and $y_3$ is larger. We find:
\begin{equation}
\begin{array}{lcrcl}
{\rm for~~} m_C^2<m_Am_D \:& (\Rightarrow& m_Am_C^2<m_B^2m_D), & \: & 
 y_1>y_3 {\rm ~if~} m_2<m_{cba}<m_4, \\
{\rm for~~} m_C^2>m_Am_D \:&{\rm \&}& m_Am_C^2>m_B^2m_D, \phantom{)} & \:& 
 y_1>y_3 {\rm ~if~} m_1<m_{cba}<m_4, \\
{\rm for~~} m_C^2>m_Am_D \:&{\rm \&}& m_Am_C^2<m_B^2m_D, \phantom{)} & \:& \\
&&
\multicolumn{3}{r}{y_1>y_3 {\rm ~if~} {\rm min}(m_1,m_2) < m_{cba} 
< {\rm max}(m_1,m_2).}
\end{array}
\end{equation}

As for the simpler two-particle invariant masses, the mass-dependent relations
obtained above result in different distributions in different regions of
mass-space. We must therefore specify the mass hierarchy of these regions
before continuing. Given the ordering of eq.~(\ref{eq:ordering}), an
unambiguous division\footnote{Note that the `regions' defined here are
different from those of sects.~\ref{sect:cHigh} and \ref{sect:cLow}. The
present ones are three-dimensional volumes, whereas those of
sects.~\ref{sect:cHigh} and \ref{sect:cLow} are two-dimensional areas.} is:
\begin{alignat}{4}
&\text{\bf Region~1:}   
&\quad  
&\underline{m_{C}^{2}<m_{A}m_{D}}, 
&\quad  
& m_{A}m_{C}^{2} < m_B^2m_D, 
&\quad  
& m_B^2 < m_A m_D
\label{eq:m_cba-Case-1} \\
&\text{\bf Region~2:}  
&\quad  
&m_{C}^{2}>m_{A}m_{D}, 
&\quad  
& \underline{m_{A}m_{C}^{2} > m_B^2m_D}, 
&\quad  
& m_B^2 < m_A m_D
\label{eq:m_cba-Case-2} \\
&\text{\bf Region~3:} 
&\quad  
&m_{C}^{2}>m_{A}m_{D}, 
&\quad  
& m_{A}m_{C}^{2} < m_B^2m_D, 
&\quad  
& \underline{m_B^2 > m_A m_D}
\label{eq:m_cba-Case-3} \\
&\text{\bf Region~4:}  
&\quad  
&m_{C}^{2}>m_{A}m_{D}, 
&\quad  
& m_{A}m_{C}^{2} < m_B^2m_D, 
&\quad  
& m_B^2 < m_A m_D
\label{eq:m_cba-Case-4}
\end{alignat}
This division, in terms of the mass ratios of eq.~(\ref{eq:ratios-R}),
is illustrated in figure~\ref{fig:limits3D}, and has the property that
each region has a different endpoint for the invariant mass
distribution.  The different orderings of $m_{1}$, $m_{2}$ and
$m_{3}$, as given by eq.~(\ref{eq:m_order}), will lead to further
subdivisions of the above regions.

\FIGURE[ht]{
\epsfxsize 15.5cm
\epsfbox{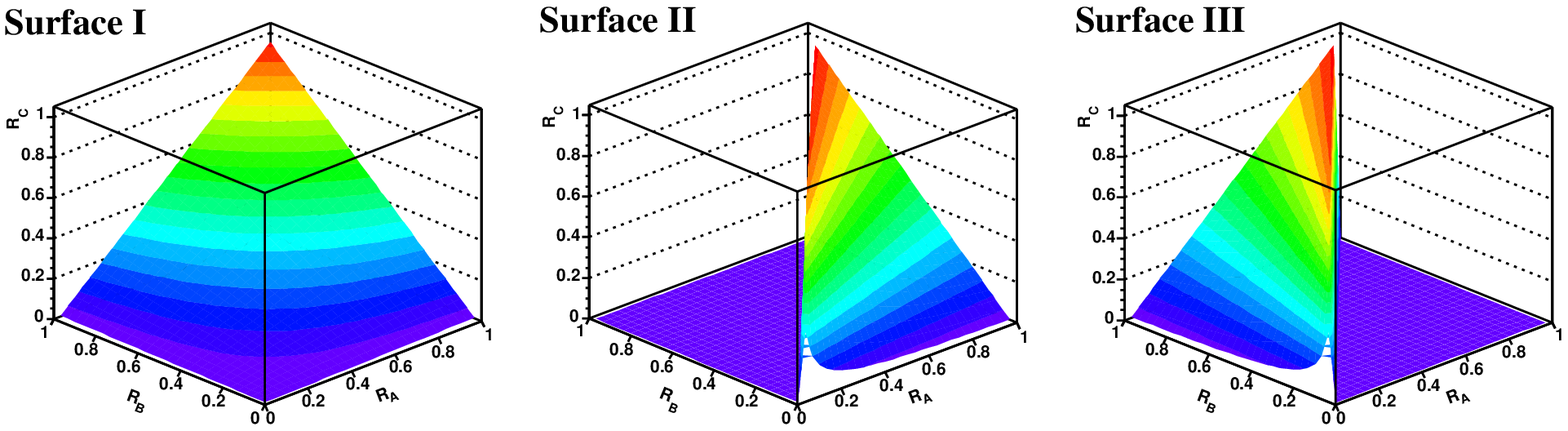}
\caption{Allowed values of $R_A$, $R_B$ and $R_C$ for the four main regions of
$m_{cba}$ in terms of volumes inside the unit cube. The allowed values for
Region~1 lie in the volume below Surface~I: $0<R_C<R_AR_B<1$. For Region~2 the
allowed volume is that above Surface~II: $0<\frac{R_B}{R_A}<R_C<1$ (not
including the flat region where $R_B>R_A$). Likewise the allowed volume for
Region~3 is that above Surface~III: $0<\frac{R_A}{R_B}<R_C<1$. Region~4 lies
in the volume above Surface~I and below Surfaces~II and III. (See
figure~\ref{fig:feetcba} for projections onto the $R_A$--$R_B$-plane).
\label{fig:limits3D}}
}

\subsubsection{Region~1: $0<R_C<R_AR_B<1$}

The first inequality (underlined) of eq.~(\ref{eq:m_cba-Case-1})
implies the other two hold for our mass hierarchy. Also, as seen from
(\ref{eq:m_4-order}) and (\ref{eq:m_order}), since
$m_{C}^{2}<m_{A}m_{D}<m_{B}m_{D}$ and
$m_{B}^{2}m_{C}^{2}<m_{C}^{4}<m_{A}^{2}m_{D}^{2}$, we have
$m_2<m_1<m_4$ and $m_3<m_1<m_4$ respectively, but the relative size of
$m_{2}$ and $m_{3}$ is undetermined.
Thus, we must divide the region into subregions.

\ssp

\noindent \underline{Region~(1,1)}: If $m_B^2>m_Am_C$ then
$m_{2}<m_{3}<m_{1}$ (see eq.~(\ref{eq:m_order})), and we find the
invariant mass distribution is, in terms of the function $L$ defined
in eq.~(\ref{eq:define-L}),
\begin{equation}
\frac{1}{\Gamma_0} \frac{\partial \Gamma_0}{\partial m_{cba}^2}
=\frac{1}{(m_{ca}^{\max})^{2}a}
\begin{cases}
L(y_{3},y_{2}) \quad
&\text{for~~} 0<m_{cba}<m_{2}, \\[4pt]
L(y_{1},y_{2}) \quad
&\text{for~~} m_{2}<m_{cba}<m_{3}, \\[4pt]
L(y_{1},y_{4}) \quad
&\text{for~~} m_{3}<m_{cba}<m_{1},
\end{cases}
\label{eq:region11}
\end{equation}
and zero otherwise.

\noindent \underline{Region~(1,2)}: Alternatively, if $m_B^2<m_Am_C$
then $m_{3}<m_{2}<m_{1}$ and the invariant mass
distribution is:
\begin{equation}
\frac{1}{\Gamma_0} \frac{\partial \Gamma_0}{\partial m_{cba}^2}
=\frac{1}{(m_{ca}^{\max})^{2}a}
\begin{cases}
L(y_{3},y_{2}) \quad
&\text{for~~} 0<m_{cba}<m_{3}, \\[4pt]
L(y_{3},y_{4}) \quad
&\text{for~~} m_{3}<m_{cba}<m_{2}, \\[4pt]
L(y_{1},y_{4}) \quad
&\text{for~~} m_{2}<m_{cba}<m_{1},
\end{cases}
\end{equation}
and zero otherwise.

\subsubsection{Region~2: $0<\frac{R_B}{R_A}<R_C<1$}

In this region, the second inequality (underlined) of
eq.~(\ref{eq:m_cba-Case-2}) implies the other two. Also, since
$m_{B}m_{D}<m_{C}^{2}$ and
$m_{B}^{2}<\frac{m_{D}}{m_{C}}m_{B}^{2}<m_{A}m_{C}$ we have
$m_{1}<m_{2}$ and $m_{3}<m_{2}$, but the relative magnitude of $m_{1}$
and $m_{3}$ is undetermined and we must divide our region in two.

\ssp

\noindent \underline{Region~(2,1)}: If $m_Am_D<m_Bm_C$ then
$m_{1}<m_{3}<m_{2}$ (see eq.~(\ref{eq:m_order})), and we find the
invariant mass distribution is:
\begin{equation}
\frac{1}{\Gamma_0} \frac{\partial \Gamma_0}{\partial m_{cba}^2}
=\frac{1}{(m_{ca}^{\max})^{2}a}
\begin{cases}
L(y_{3},y_{2}) \quad
&\text{for~~} 0<m_{cba}<m_{1}, \\[4pt]
L(y_{1},y_{2}) \quad
&\text{for~~} m_{1}<m_{cba}<m_{3}, \\[4pt]
L(y_{1},y_{4}) \quad
&\text{for~~} m_{3}<m_{cba}<m_{2},
\end{cases}
\end{equation}
and zero otherwise.

\ssp

\noindent \underline{Region~(2,2)}: If $m_Am_D>m_Bm_C$ then
$m_{3}<m_{1}<m_{2}$, and the distribution is:
\begin{equation}
\frac{1}{\Gamma_0} \frac{\partial \Gamma_0}{\partial m_{cba}^2}
=\frac{1}{(m_{ca}^{\max})^{2}a}
\begin{cases}
L(y_{3},y_{2}) \quad
&\text{for~~} 0<m_{cba}<m_{3}, \\[4pt]
L(y_{3},y_{4}) \quad
&\text{for~~} m_{3}<m_{cba}<m_{1}, \\[4pt]
L(y_{1},y_{4}) \quad
&\text{for~~} m_{1}<m_{cba}<m_{2},
\end{cases}
\end{equation}
and zero otherwise.

\subsubsection{Region~3: $0<\frac{R_A}{R_B}<R_C<1$}

In this region, it is the third inequality (underlined) which leads to
the other two. The inequalities $m_{A}m_{C}<m_{A}m_{D}<m_{B}^{2}$ and
$m_{A}m_{D}<m_{B}m_{C}$ tell us that $m_{2}<m_{3}$ and
$m_{1}<m_{3}$. Although we do not know the order of $m_1$ and $m_2$,
the endpoints of the integration are independent of which one is
larger and their relative magnitude only affects the region of
applicability of the different functions
(unlike in Regions 1 and 2 where the integration
endpoints also changed). For the entirety of Region 3 we find:
\begin{equation}
\frac{1}{\Gamma_0} \frac{\partial \Gamma_0}{\partial m_{cba}^2} =
\frac{1}{(m_{ca}^{\max})^{2}a}
\begin{cases}
L(y_{3},y_{2}) \quad
&\text{for~~} 0<m_{cba}<\min(m_1,m_2), \\[4pt]
L(y_{1},y_{2}) \quad
&\text{for~~} \min(m_1,m_2)<m_{cba}<\max(m_1,m_2), \\[4pt]
L(y_{3},y_{2}) \quad
&\text{for~~} \max(m_1,m_2)<m_{cba}<m_{3},
\end{cases}
\end{equation}
and zero otherwise. Note that $m_1<m_2$ when $m_Bm_D<m_C^2$, from
eq.~(\ref{eq:m_order}).

\subsubsection{Region~4: $0<R_AR_B<R_C$ and $R_C<\frac{R_B}{R_A}<1$
and $R_C<\frac{R_A}{R_B}<1$}

Region 4 encompasses all the other allowed areas of mass-space. In this region
we only know, {\it a priori}, that $m_1$, $m_2$ and $m_3$ are smaller than
$m_4$ (as for all regions) but not their relative sizes. However, as for
Region~3, the ordering of $m_1$ and $m_2$ does not change the endpoints of the
integration, but only changes the region of applicability. Therefore, we have
three possible subregions, corresonding to the relative sizes of $m_3$ and the
maximum and minimum of $m_1$ and $m_2$.

\ssp

\noindent \underline{Region~(4,1)}: If $m_Am_D>m_Bm_C$ and
$m_B^2<m_Am_C$ then $m_3<\min(m_1,m_{2})$, and the invariant mass
distribution is:
\begin{equation}
\frac{1}{\Gamma_0} \frac{\partial \Gamma_0}{\partial m_{cba}^2}=
\frac{1}{(m_{ca}^{\max})^{2}a}
\begin{cases}
L(y_{3},y_{2}) \quad
&\text{for~~} 0<m_{cba}<m_{3}, \\[4pt]
L(y_{3},y_{4}) \quad
&\text{for~~} m_{3}<m_{cba}<\min(m_1,m_2), \\[4pt]
L(y_{1},y_{4}) \quad
&\text{for~~} \min(m_1,m_2)<m_{cba}<\max(m_{1},m_2),\\[4pt]
L(y_{3},y_{4}) \quad
&\text{for~~} \max(m_1,m_2)<m_{cba}<m_{4},
\end{cases}
\end{equation}
and zero otherwise.

\ssp

\noindent \underline{Region~(4,2)}: If \{$m_Bm_D>m_C^2$, $m_Am_D>m_Bm_C$
and $m_B^2>m_Am_C$\} {\bf or} \{$m_Bm_D<m_C^2$, $m_Am_D<m_Bm_C$ and
$m_B^2<m_Am_C$\} then $\min(m_1,m_{2})<m_3<\max(m_1,m_2)$, and the
invariant mass distribution is:
\begin{equation}
\frac{1}{\Gamma_0} \frac{\partial \Gamma_0}{\partial m_{cba}^2}=
\frac{1}{(m_{ca}^{\max})^{2}a}
\begin{cases}
L(y_{3},y_{2}) \quad
&\text{for~~} 0<m_{cba}<\min(m_1,m_2), \\[4pt]
L(y_{1},y_{2}) \quad
&\text{for~~} \min(m_1,m_2)<m_{cba}<m_3, \\[4pt]
L(y_{1},y_{4}) \quad
&\text{for~~} m_3<m_{cba}<\max(m_1,m_2),\\[4pt]
L(y_{3},y_{4}) \quad
&\text{for~~} \max(m_1,m_2)<m_{cba}<m_{4},
\end{cases}
\end{equation}
and zero otherwise.

\ssp

\noindent \underline{Region~(4,3)}: Finally, if $m_Am_D<m_Bm_C$ and
$m_B^2>m_Am_C$ then $\max(m_1,m_{2})<m_3$, and the invariant mass
distribution is:
\begin{equation}
\frac{1}{\Gamma_0} \frac{\partial \Gamma_0}{\partial m_{cba}^2}=
\frac{1}{(m_{ca}^{\max})^{2}a}
\begin{cases}
L(y_{3},y_{2}) \quad
&\text{for~~} 0<m_{cba}<\min(m_1,m_2), \\[4pt]
L(y_{1},y_{2}) \quad
&\text{for~~} \min(m_1,m_2)<m_{cba}<\max(m_1,m_2), \\[4pt]
L(y_{3},y_{2}) \quad
&\text{for~~} \max(m_1,m_2)<m_{cba}<m_3,\\[4pt]
L(y_{3},y_{4}) \quad
&\text{for~~} m_3<m_{cba}<m_{4},
\end{cases}
\label{eq:region43}
\end{equation}
and zero otherwise.

\section{Parton level}
\label{sect:parton}

In order to explore how the derived expressions can be applied to real
data, we will compare with SUSY cascade decays in Monte Carlo (MC)
events generated for the mSUGRA Snowmass benchmark point
SPS~1a~\cite{Allanach:2002nj}.  SPS~1a has the GUT scale parameters
$m_{0}=100$~GeV, $m_{1/2}=250$~GeV, $A_{0}=-100$~GeV, $\tan\beta=10$
and $\mu>0$.  The TeV-scale SUSY mass spectrum is calculated from
these parameters, together with a top mass of $m_{t}=175$~GeV, by
ISAJET~7.58 \cite{Baer:1993ae} (for numerical values, see, e.g.\
\cite{Gjelsten:2004}). We have generated a number of events equivalent
to $50~\text{fb}^{-1}$ at the LHC, by running PYTHIA~6.208
\cite{PYTHIA} with CTEQ~5L parton distribution functions
\cite{Lai:1999wy}.

In particular, we focus on the decay chain (\ref{eq:squarkchain}),
which was previously investigated in some detail in
\cite{Gjelsten:2004}. The comparison is done for the invariant masses
$\mqlLow$, $\mqlHigh$, $m_{ql_{f}}$ and $m_{qll}$.\footnote{We include
the distribution of $m_{ql_{f}}$ for comparison with our analytical
expression, even though it will be difficult to separate the near and
far leptons experimentally.} For the quark $q$ we consider only the up
quark and the corresponding squark. The complications introduced with
more than one flavour, and thus multiple, overlaying distributions,
will be faced in the next section.

We find the analytic expressions for the distributions by going from
the distribution of the square of the invariant mass to the
distribution of the invariant mass by eq.~(\ref{eq:m2tom}) and
substituting $A=\tilde{\chi}^{0}_{1}$, $B=\tilde{l}_{R}$,
$C=\tilde{\chi}^{0}_{2}$ and $D=\tilde{u}_{L}$ in our
expressions.

However, this simple picture is not what will be seen in an actual
experiment. The shapes will be distorted by the width of the SUSY
particles, cuts applied to remove backgrounds, by Final State
Radiation (FSR), and by detector effects.  In order to use the shapes for
extracting NP particle masses from real data it is important to
understand these effects, and the limits they place on the use of the
shapes.  Here we consider the three main effects apparent at parton
level, width effects, the bias introduced by cuts and a shift in
invariant mass from FSR.

\subsection{Width}
Since we have not taken into account the width of the
decaying particles we additionally smear the analytic expression with
a (truncated) Breit-Wigner function. If $g(m)$ is the original
distribution we plot in the following figures the function $f(m)$
given by the convolution
\begin{equation}
f(m)=\frac{\Gamma_s}{4\arctan 20}\int_{m-10\Gamma_s}^{m+10\Gamma_s}
\frac{g(m')}{(m'-m)^2+(\Gamma_s/2)^2}dm',
\end{equation}
where $\Gamma_s$ is a parameter of the smearing,
determined by a fit to the MC data. We normalize the
analytic expression to the number of MC events.  

The smeared analytic curves (for the SPS~1a parameters) are compared
with the parton level MC distributions, with no cuts or Final State
Radiation, in figure~\ref{fig:nocuts}.
\FIGURE[t]{
\epsfxsize 13cm
\epsfbox{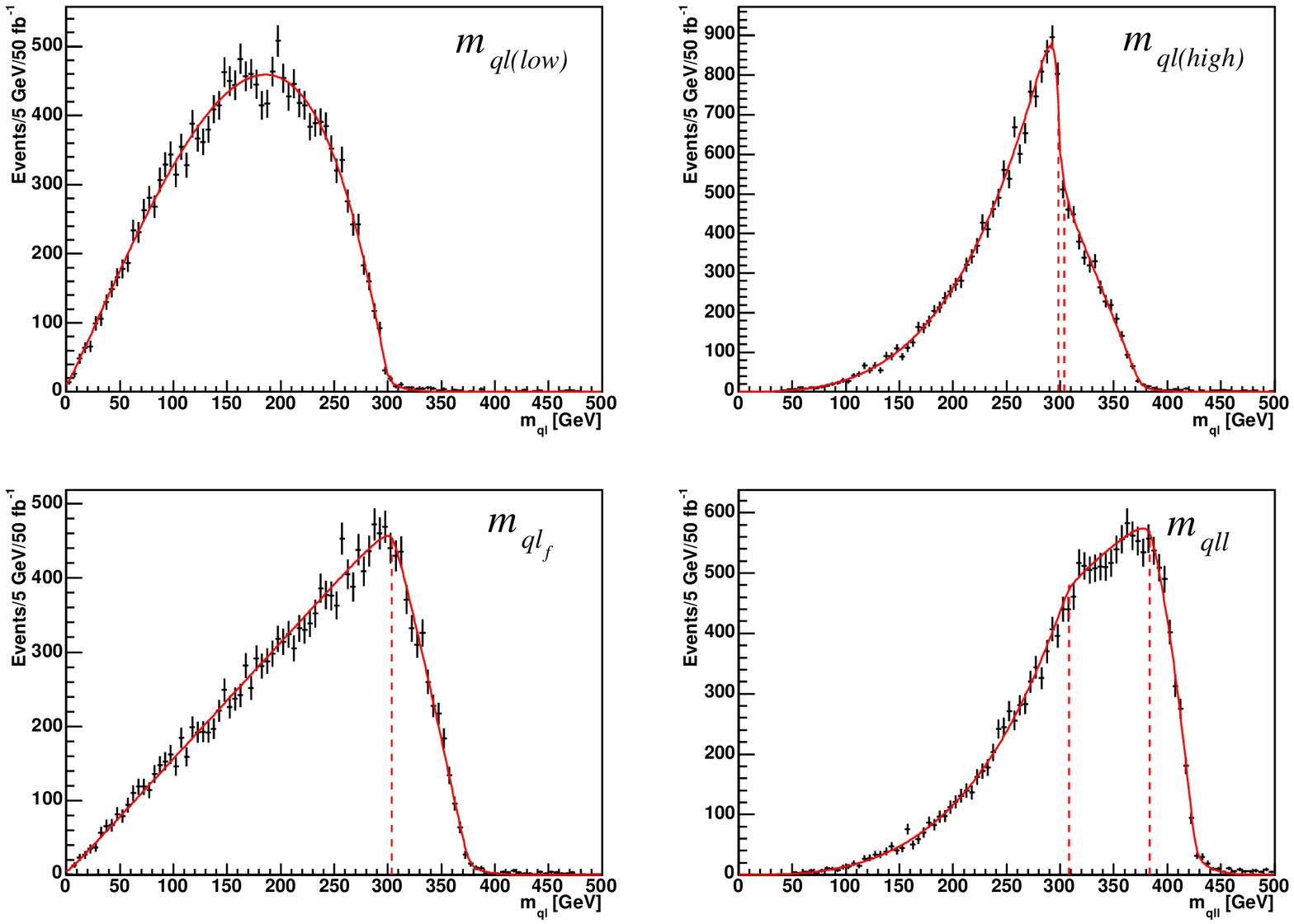}
\caption{Invariant mass distributions at parton level with no cuts and no
FSR. MC events (black) are shown with errors. The lines (red) are the
analytic expressions with nominal masses. The dashed lines show the
ranges of validity of the different function pieces.
\label{fig:nocuts}}
}
The agreement is very good, being at the level expected from
statistical fluctuations in the MC, and provides a verification of our
analytic expressions. The smearing is responisble for a rounding of
the sharp peaks in the $\mqlLow$ and $m_{cba}$ distributions, and
greatly improves agreement near the upper end of the distributions.

\subsection{Cuts} \label{sect:cuts}

We apply the cuts used in \cite{Gjelsten:2004}
to isolate the decay chain from Standard Model background,
\begin{itemize}
\item the three hardest jets have  $p^{\text{jet}}_T>150,100,50~\text{GeV}$,
\item $E_{T,\text{miss}}>\max(100~\text{GeV}, 0.2~M_\text{eff})$, where
$M_\text{eff}\equiv E_{T,\text{miss}}+\sum^3_{i=1} p^{\text{jet}}_{T,i}$,
\item the two hardest leptons have $p^{\text{lep}}_{T}>20,10~\text{GeV}$.
\end{itemize}
Figure~\ref{fig:partnofsr} compares the analytic curves with the parton
level MC events after the application of these cuts.
\FIGURE[ht]{
\epsfxsize 13cm
\epsfbox{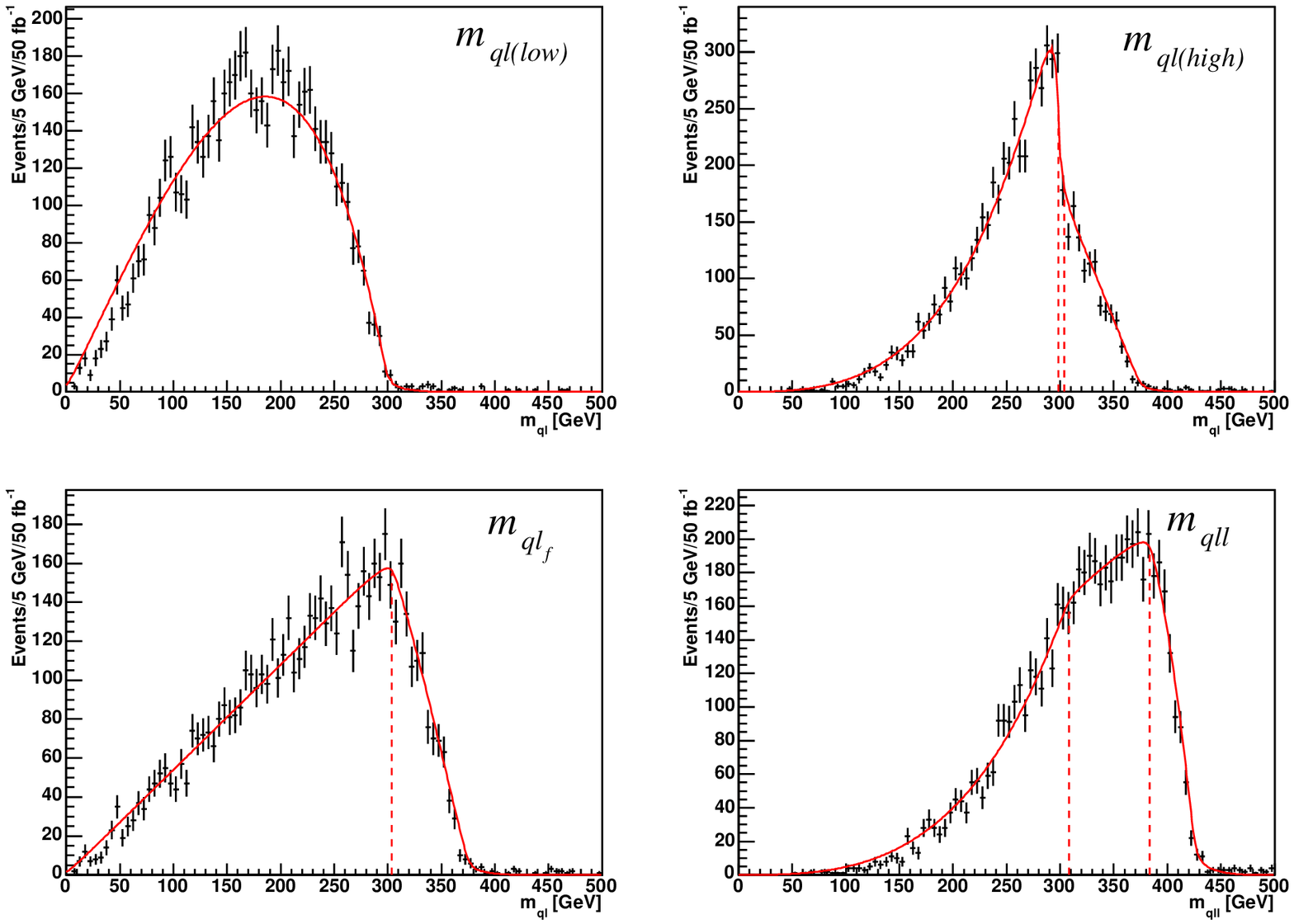}
\caption{Invariant mass distributions at parton level with cuts as described
in the text, but no FSR. MC events (black) are shown with errors. The lines
(red) are the analytic expressions with nominal masses. The dashed lines show
the ranges of validity of the different function pieces.
\label{fig:partnofsr}}
}
From a visual inspection we see a fairly good agreement between the MC
events and the analytic functions, except that the number of events at
low invariant mass is reduced in the MC distribution
compared to the analytic curve. This is most pronounced for $\mqlLow$.

In order to determine which of the cuts is introducing this
discrepancy, we show in figure~\ref{fig:cutavwcuts} the average value of the
cut variables $E_{T,\text{miss}}-\max(100,0.2M_{\text{eff}})$,
$p^{\text{lep}}_{T,1}$, $p^{\text{lep}}_{T,2}$ and
$p^{\text{jet}}_{T,1}$, in each bin of the $\mqlLow$ invariant mass
distribution versus the invariant mass in that bin. The two jet cuts
omitted have a very similar behaviour to that on the hardest jet.
\FIGURE[ht]{
\epsfxsize 13cm
\epsfbox{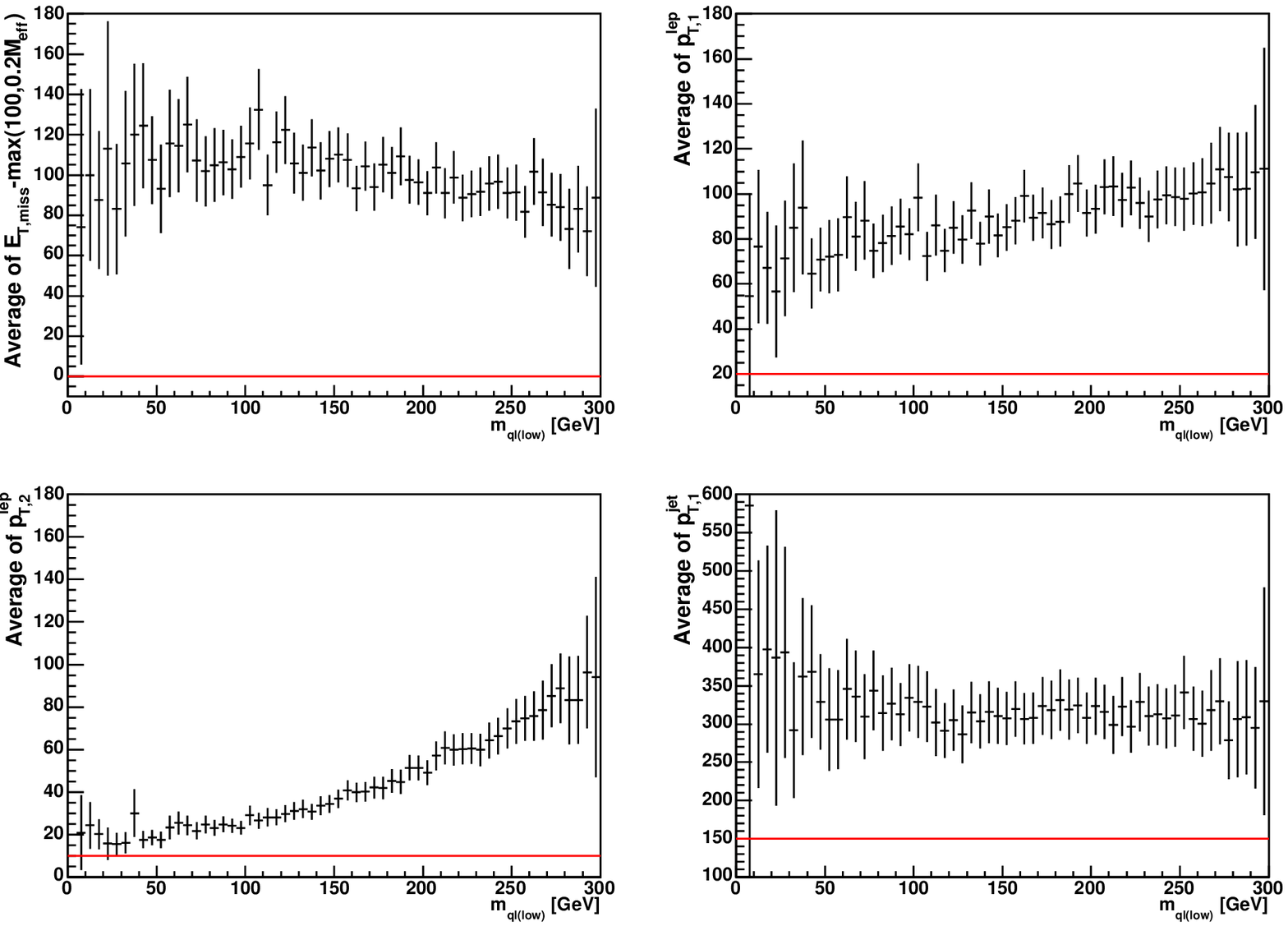}
\caption{Average value of the cut variables
$E_{T,\text{miss}}-\max(100,0.2M_{\text{eff}})$, $p^{\text{lep}}_{T,1}$,
$p^{\text{lep}}_{T,2}$ and $p^{\text{jet}}_{T,1}$, versus invariant mass for
$\mqlLow$ (black). We impose the cuts described in the text on the invariant
mass distribution before computing averages, but include no FSR. 
The horizontal lines (red) show the cuts described in the text.
\label{fig:cutavwcuts}}
}

For the cut on missing energy there is a large gap between the signals's
missing energy and the cut value (except for the two end bins, where low
statistics lead to large errors) and therefore the effects on the shape of the
invariant mass distribution will be small. On the other hand, because of the
small difference between the cut value and the average value at low invariant
masses, the cut on the transverse momentum of the second hardest lepton,
$p^{\text{lep}}_{T,2}$, will clearly introduce a bias by removing more events
at low invariant masses. From figure~\ref{fig:cutavwcuts} we see that we
should be safe in trusting the analytic shape of the distribution after the
cut on $p^{\text{lep}}_{T,2}$, down to around $\mqlLow=100-150$ GeV where the
$p^{\text{lep}}_{T,2}$ distribution flattens out due to the cut imposed. The
cut on the transverse momentum of the hardest lepton, $p^{\text{lep}}_{T,1}$,
will likewise have some effect on the distribution, but only at very low
invariant masses. For the cut on transverse momentum of the hardest jet, and
similarly for the other two jet cuts, we again find that the average values do
not lie close enough to the cut value, for most of the invariant mass bins, to
cause a problem. However, even if this had not been the case, since the
average value of $p^{\text{jet}}_{T,1}$ as a function of $\mqlLow$ is to a
good approximation flat, it would have cut equally at all invariant mass
values, and so introduced no bias.

Studying the average of a cut variable over the range of possible
invariant mass values enables us to give an estimate of where it
is safe to fit the distributions given some cut value,
and could also be used to optimize the cuts so that
they introduce less or no bias for certain ranges of invariant mass.

\subsection{Final state radiation}

In figure~\ref{fig:part} we include FSR in the MC data
and we observe a slight shift towards lower invariant masses for all
four distributions as compared to the analytic shape.  This can be
compared with figure~\ref{fig:partnofsr} which included no FSR.

\FIGURE[ht]{
\epsfxsize 13cm
\epsfbox{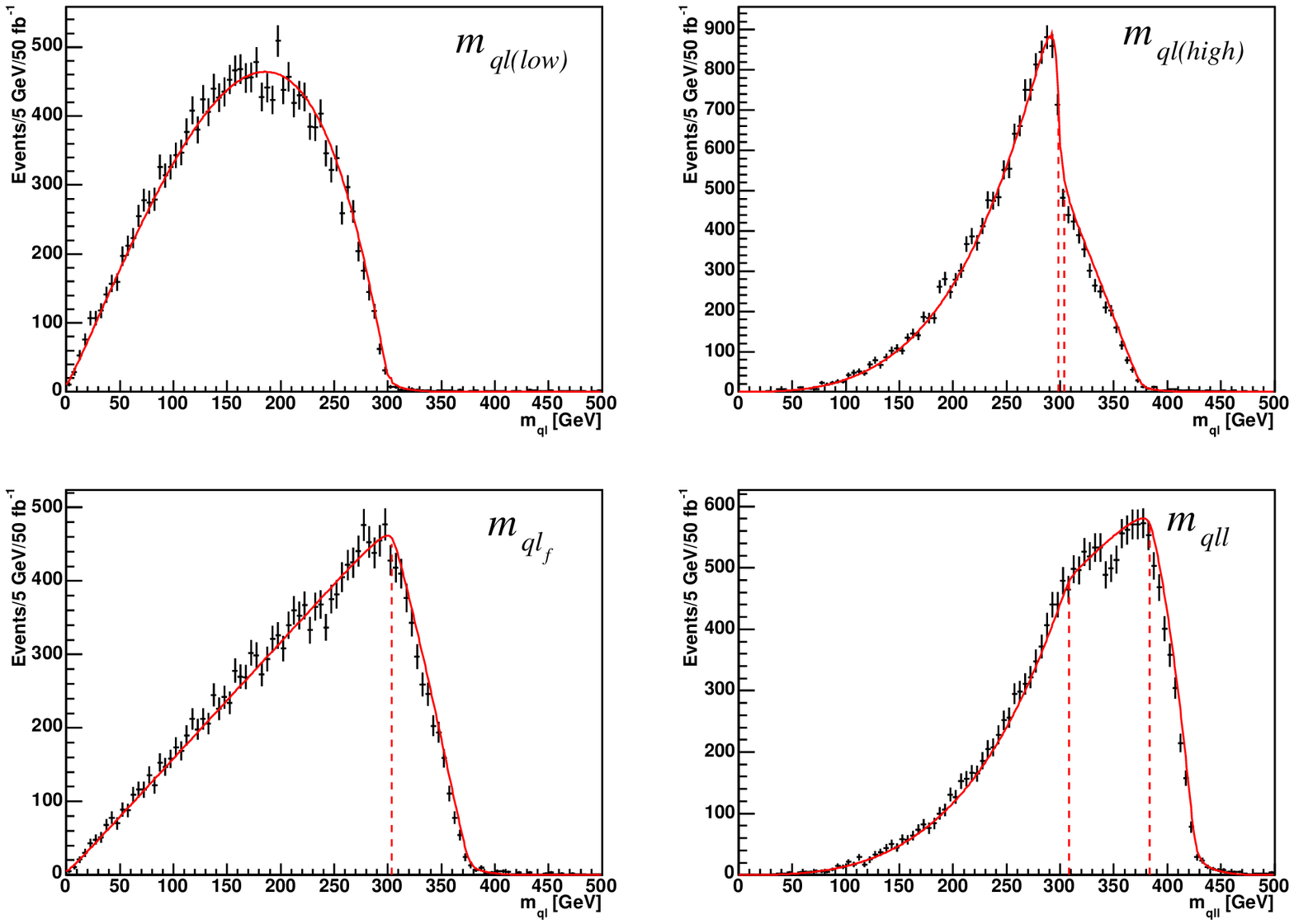}
\caption{Invariant mass distributions at parton level with FSR included,
but no cuts.
MC events are shown with errors (black). The lines (red) are the
analytic expression with nominal masses. The dashed lines show the
ranges of validity of the different function pieces.
\label{fig:part}}
}

Ignoring statistical fluctuations between the two different event samples
this shows that FSR conserves the shape of the distributions, but
lowers the invariant masses slightly, indicating that it should be
possible to use the shape of the distributions in
determining SUSY masses, but that the precision will be limited by
that of the jet energy scale.

\section{Detector effects}
\label{sect:detector}

In section~\ref{sect:parton} we showed that the shape of the invariant mass
distributions from the decay chain in eq.~(\ref{eq:squarkchain}) to a large
extent survived parton level effects and cuts to remove SM background.  We
demonstrated a method of showing whether and where cuts could deform the
distribution, making our shape predictions from the kinematics of the decay
chain unreliable. What remains to be discussed are the effects of the detector
in a given experiment and the combinatorical complications introduced by
trying to pick jets that correspond to the quarks in the decay chain.

For this purpose we use AcerDET 1.0 \cite{Richter-Was:2002ch}, a
generic fast detector simulation for the LHC, similar in structure to
the ATLFAST \cite{ATLFAST} MC simulation of the ATLAS
detector. AcerDET expresses identification and isolation of leptons
and jets in terms of detector coordinates by azimuthal angle $\phi$,
pseudo-rapidity $\eta$ and cone size $\Delta
R=\sqrt{(\Delta\phi)^2+(\Delta\eta)^2}$. We identify a lepton if
$p_T>5(6)$~GeV and $|\eta|<2.5$ for electrons (muons). A lepton is
isolated if it is a distance $R>0.4$ from other leptons and jets and
the transverse energy deposited in a cone $\Delta R=0.2$ around the lepton
is less than $10$~GeV. Jets are reconstructed by a cone-based algoritm
from clusters and are accepted if the jet has $p_T>15$~GeV in a cone
$\Delta R=0.4$. The jets are recalibrated using a flavour independent
parametrization, optimized to give a proper scale for the dijet decay
of a light ($100$~GeV) Higgs particle.

As in the parton level discussion we use the mSUGRA model point
SPS~1a, a number of events equivalent to $50~\text{fb}^{-1}$ of
integrated luminosity generated by PYTHIA~6.208 \cite{PYTHIA}, and the
cuts of subsect.~\ref{sect:cuts}. In addition to the cuts introduced
earlier we also cut on $b$-tagged jets to remove events with a
$b$-squark in the decay chain, which will, due to a smaller squark
mass, have a different distribution.  We assume $50\%$ $b$-tagging
efficiency with rejection factors 100 and 10 on jets from gluons/three
lightest jets and $c$-jets respectivly.  While this is on the
conservative side it removes a majority of events with $b$-squarks in
the decay chain. However some events remain that will subtly change
the distribution. How large a problem this will be will depend on the
rate of $\tilde b$ production and the efficiency achieved for
$b$-tagging. We have the same issue for the $u$ and $d$-squarks, since
they in general have different masses, but here we cannot tag the
jets. One possible solution to this problem is to fit the experimental
distributions with the weighted sum of two functions with different
squark mass values. This effect is of course smaller than for the
$b$-squark, as the mass difference of these squarks is relatively
small in our scenario.

Even if we could remove effectively all events with $b$-jets another
potentially large problem remains. We have no sure way of knowing
which jet corresponds to the quark in our decay chain. Since the other
jets can stem from the decay of the other SUSY particle
produced in the hard process or from the underlying event we cannot
rely on ``high-low'' distributions as is done with the two leptons. We
therefore propose to use consistency cuts, as discussed in
\cite{Gjelsten:2004}, to purify the events.  We assume that the
endpoints of the distributions have already been estimated, but not
necessarily very precisely. To plot the distribution of a given
invariant mass among the set: $\mqlLow$, $\mqlHigh$ and
$m_{qll}$, we then cut away all events except those where one and only
one of the two hardest jets, when combined with the leptons, gives
invariant masses that lie below both of the endpoints of the
distributions we are not plotting, i.e.\ that the jet we pick for an
event in a given distribution is consistent with the other
distributions and that there is no other such consistent jet, among
the two hardest ones.  Additionally we require that the two leptons
have an invariant mass below the endpoint of the $m_{ll}$ distribution.
As we shall see, these consistency cuts are very effective in leaving
only events where we have picked the right jet to go with the leptons.
However they will reduce the number of accepted events significantly,
and thus increase statistical errors.

The distributions of the invariant masses $\mqlLow$, $\mqlHigh$ and
$m_{qll}$, after detector simulation, after the cuts of
section~\ref{sect:cuts} and after cuts on $b$-tagged jets and
consistency cuts, are shown in figure~\ref{fig:deteff} with error bars
(black). For the consistency cuts we have used as endpoint values
$\mqlLow^{\max}=320$~GeV, $\mqlHigh^{\max}=395$~GeV and
$m_{qll}^{\max}=450$~GeV.\footnote{For $m_{ll}^{\max}$ we use $80$~GeV
which is $\sim3$~GeV above the fit value.}  These lie $\sim20$~GeV
above the fit values of \cite{Gjelsten:2004}, and are clearly well
outside the level of uncertainty expected for the measurement of the
endpoints, showing that the consistency cuts will not critically
depend on measuring the endpoints very accurately.  Also in
figure~\ref{fig:deteff} (blue histograms) we show the parton level
distributions after cuts, rescaled to the reduced number of events in
the detector level distributions. Finally, we show the nominal
distributions, as given from our formulae (red curves).

\FIGURE[ht]{
\epsfxsize 15cm
\epsfbox{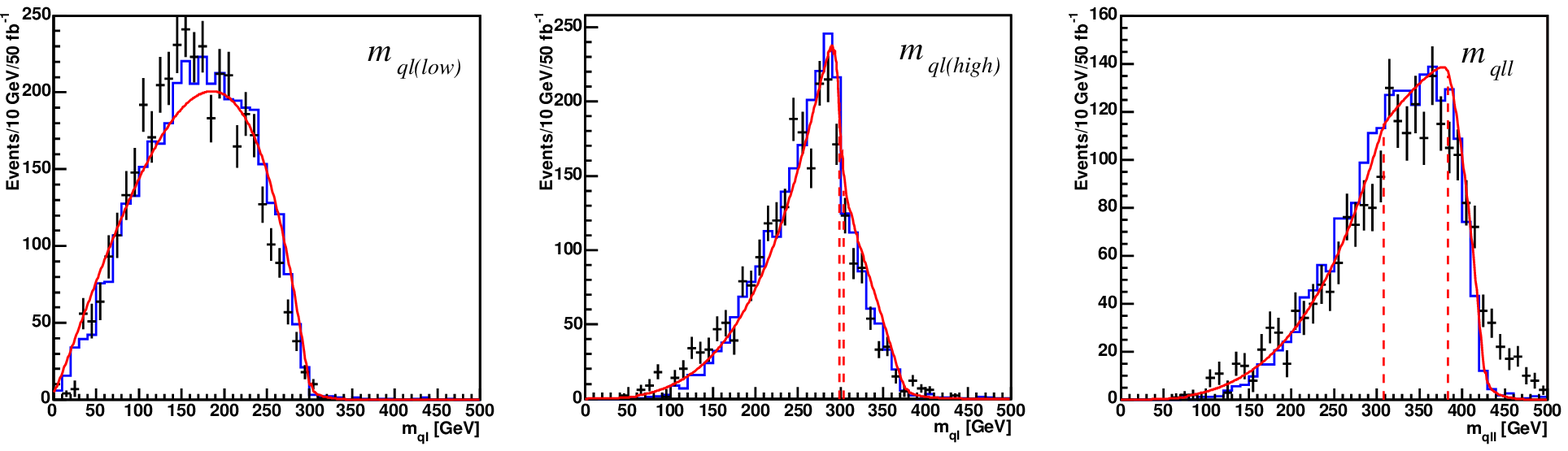}
\caption{Invariant mass distributions after detector effects and consistency
cuts shown with error bars (black).  Also shown, are the parton level
distributions (blue histograms) with the same normalization as the
detector signal and the nominal shapes of the distributions (red curves).
\label{fig:deteff}}
}

From figure~\ref{fig:deteff} we see that the distributions for
$\mqlLow$, $\mqlHigh$ and $m_{qll}$ in general remain unchanged after
detector effects, but compared to figure~\ref{fig:part} we have fewer
events and thus larger errors. In the $\mqlHigh$ distribution we see a
rounding of the peak due to additional smearing from energy
measurements in the detector simulation. All three distributions are also
shiftet slightly to the left (lower invariant masses) of the nominal
shape as a result of inperfect jet recalibration. For the
$m_{qll}$ distribution we note that some events with an erroneously
picked jet remain to the right of the nominal endpoint of the
distribution. The number of these events can be reduced by tighter
consistency cuts. Knowing the expected shape of the signal
distribution, one could also model the shape of the background and
subtract it.

\section{Feet}
\label{sect:feet}

As discussed in \cite{Gjelsten:2004}, the mass values may be such that
some invariant mass distributions exhibit ``feet'' or ``drops'' at the
high ends. These can be hidden by a significant
presence of background, taken to be smearing from detector effects or
even assumed to be a width effect of the sparticles, making a precise
determination of the kinematic maximum, and through these the SUSY
masses, difficult and subject to systematic errors.

There are two basic features in the distributions that can result in
feet: The first is that the last function piece has a maximum value
that is much lower than the global maximum, and so is not taken to be
part of the distribution. This we will refer to as a ``foot''.  The
second feature comes about when the last function piece does not fall
gradually to zero, but ends in a discontinuous jump.  This we will
call a ``drop''.  The two situations are illustrated in
figure~\ref{fig:feetill}. We will often refer collectively to ``feet''
and ``drops'' simply as ``feet''.

\FIGURE[ht]{
\epsfxsize 10cm
\epsfbox{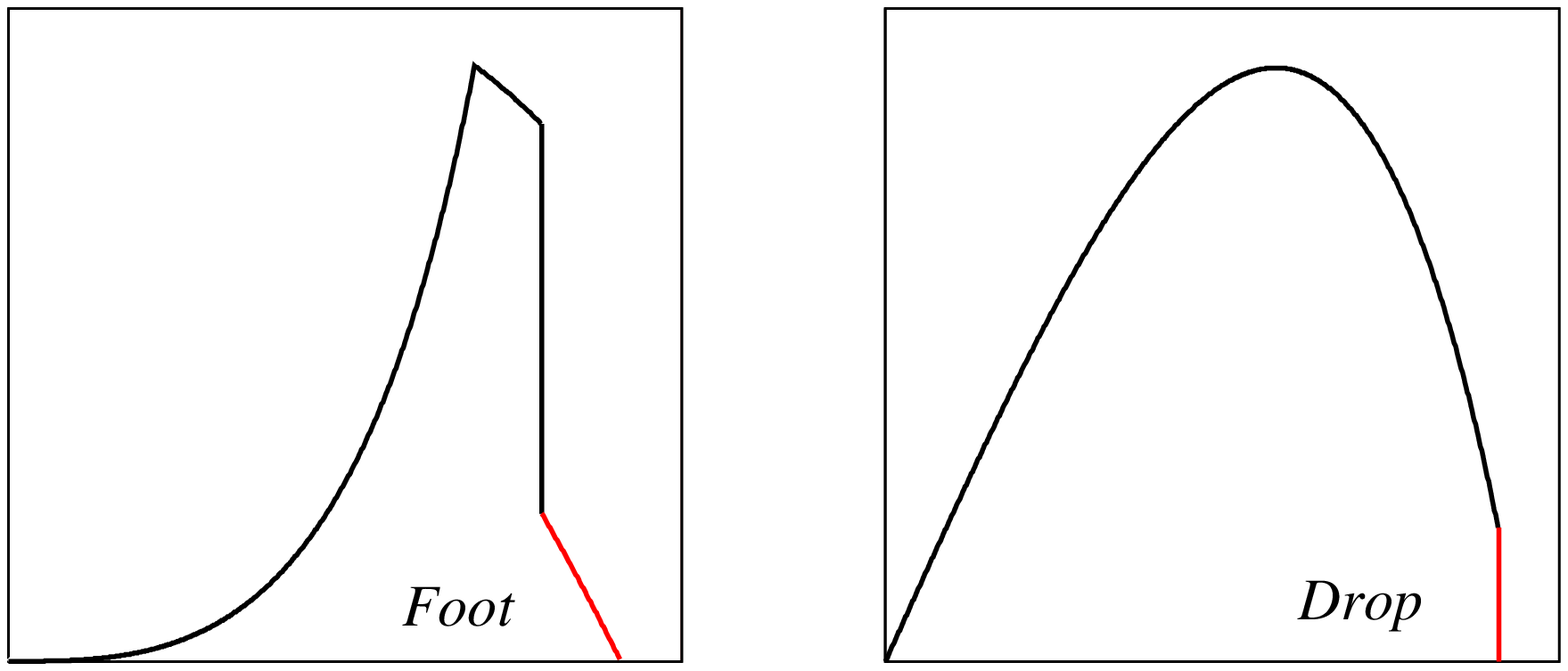}
\caption{Example of a ``foot'' and a ``drop''.
\label{fig:feetill}}
}

For both cases, a low value of the ratio between the
maximum of the last function piece or the height of the drop, and the
global maximum of the distribution will indicate the possible danger
of mismeasurements due to feet. We denote this ratio $r$:
\begin{equation}
\label{Eq:r}
r=\frac{\text{height of ``foot'' or ``drop''}}{\text{global maximum}}.
\end{equation}
Without reference to a specific model and thus the size of the background
compared to the signal, one can not give an exact value of $r$ where this
danger is real, but what can be done is to perform a model independent
exploration of what relationships between mass parameters give low ratios.

We will here discuss this for the invariant masses $m_{\cHigh}$,
$m_{\cLow}$ and $m_{cba}$, in terms of the mass-squared ratios
$R_{A}$, $R_{B}$ and $R_{C}$ introduced in (\ref{eq:ratios-R}). With
experimental data one could then use knowledge of the signal to
background ratio, with measured values of the mass-squared
ratios\footnote{The ratios $R_{A}$, $R_{B}$ and $R_{C}$ can to some
extent be determined from the shapes of the distributions alone, and
are thus not so susceptible to the effects of a mismeasurement of an
endpoint.}, to look for such mismeasurements. Finally we will look at
the situation for these invariant masses in the $m_{1/2}$--$m_0$-parameter
planes around the Snowmass mSUGRA benchmark points SPS1a, SPS1b, SPS3
and SPS5 \cite{Allanach:2002nj}.

\subsection{$m_{\cHigh}$}
We start our discussion of feet in the different mass distributions by an
overview of the {\it number} of function pieces involved in each region.  For
the two-particle masses $m_\cHigh$ and $m_\cLow$, these are given in
table~\ref{table:no-2-functions}.

\begin{TABLE}{
\begin{tabular}{|l|c|c|c|}
\hline
Distribution &Region~1 &Region~2&Region~3 \\
\hline
$m_\cHigh$ & 3 & 3 & 3 \\
$m_\cLow$  & 1 & 2 & 2 \\
\hline
\end{tabular}
\caption{Number of distinct functions for two-particle masses.}
\label{table:no-2-functions}}
\end{TABLE}

For the $m_{\cHigh}$ distribution the $m_{D}$ mass can be factored out
of all the function pieces for any value of $m_{\cHigh}$.  Taking the
ratio of two function pieces, these factors cancel, making the ratio
independent of $R_{C}$.  For the first two regions we can then look
for a possible foot by finding the ratio $r$ of the maximum of the
last function piece to the global maximum in terms of $R_{A}$ and
$R_{B}$.  The third function piece of the third region does not end at
zero, thus it may have a dangerous drop, as described above.  The
ratio $r$ of the function value at the endpoint to the global maximum
can likewise be given in terms of only $R_{A}$ and $R_{B}$.  The
$r$-values of all three regions are plotted in
figure~\ref{fig:feethighlow}.  The results largely agree with those
obtained for a particular set of mSUGRA model points in sec.~4.1 of
\cite{Gjelsten:2004}, where the shape of the distribution was
generated by a Monte Carlo decay routine.

\FIGURE[ht]{
\epsfxsize 14cm
\epsfbox{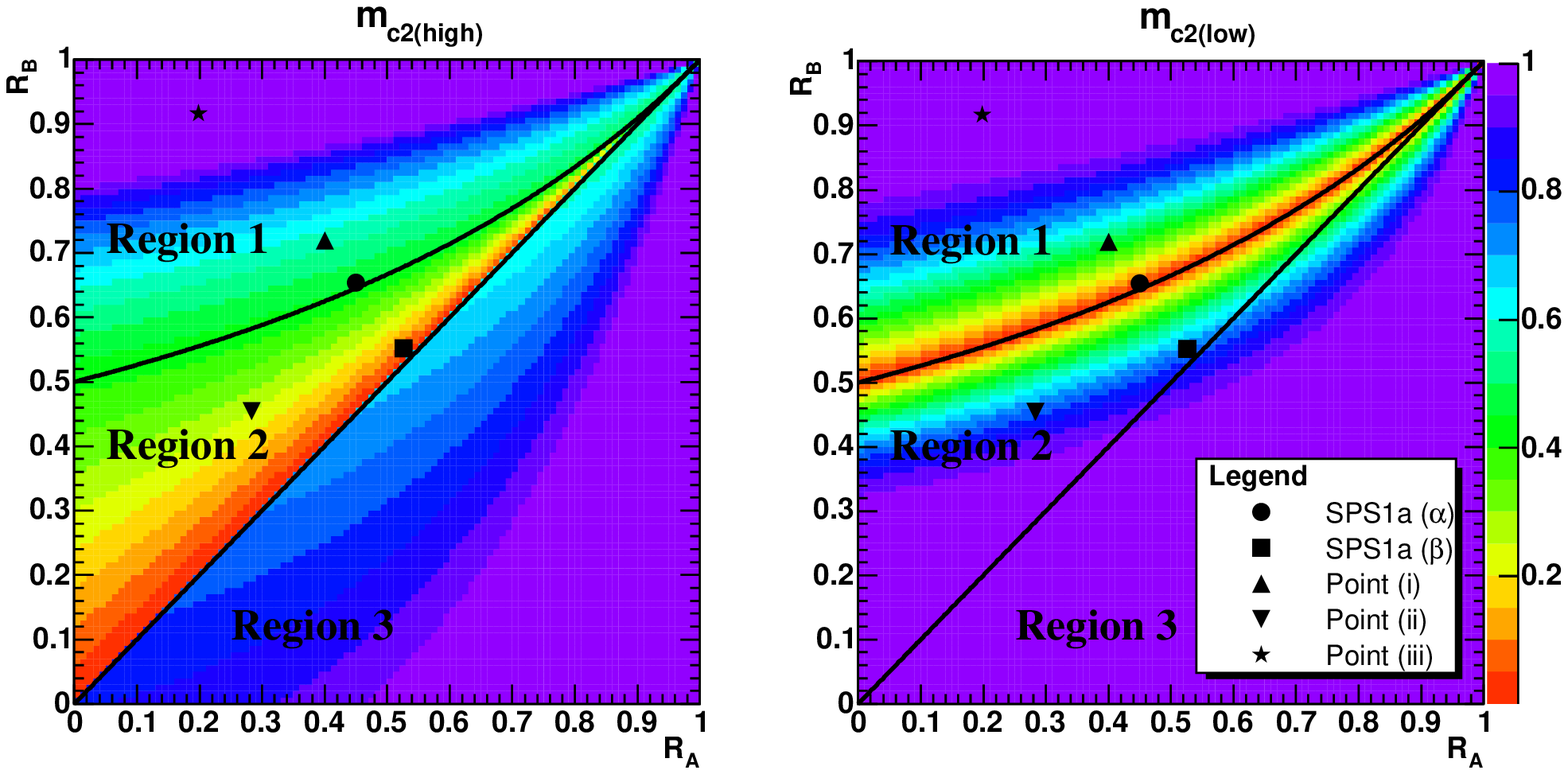}
\caption{Contour lines of $r$ for all three regions of the $m_{\cHigh}$ (left)
and $m_{\cLow}$ (right) distributions, plotted in the $R_{A}R_{B}$-plane. We
also show the position of the points SPS~1a ($\alpha$), SPS~1a ($\beta$), (i),
(ii) and (iii), referred to in the discussion of ``feet'' in
\cite{Gjelsten:2004}.
\label{fig:feethighlow}}
}

We find that for Region~1 the ratio is large for most values of $R_{A}$ and
$R_{B}$, so that the danger of a foot is small, unless we have a large
background to signal ratio.  On the other hand, in Region 2, there are large
areas with a small value of $r$ where we can have foot effects. One can show
that, as a function of $R_{A}$ and $R_{B}$, $r$ is for $m_{\cHigh}$ continuous
over the boundary between Regions~1 and 2.  Approaching the border to Region~3
the ratio goes to zero. In Region~3, $r$ again takes on large values, even
more so than for Region~1.  A minimum around $r\simeq0.60$ excludes the drop
scenario from being a danger in this region.

\subsection{$m_{\cLow}$}
For the $m_{\cLow}$ distribution we can also factor out $m_{D}$ in all
regions, with the same factor for all function pieces, so that $r$
again is independent of $m_{D}$.  In Region~1 the only function piece
does not end at zero, and we may thus encounter the drop situation.
In Regions~2 and 3 we have a potential foot from the second function
piece. In figure~\ref{fig:feethighlow}, we plot the ratio $r$ of 
eq.~(\ref{Eq:r}).

In Region~1 we find a strip along the border to Region~2 where
we can have a dangerous drop in the distribution.
Indeed SPS~1a lies in this region, which is somewhat surprising given
that this feature was not noticed in \cite{Gjelsten:2004}.
This clearly shows the advantage of having analytical expressions
for the shapes of the distributions.
However, the drop here is so small, $r\approx0.046$,
that the resulting mismeasurement of the endpoint amounts
to $0.1\%$, much less than the statistical error.
For Regions~2 and 3 $r$ is again continuous over the border.
The potential for a foot is however restricted to a small strip along
the border to Region~1.

\subsection{$m_{cba}$}
The numbers of function pieces found in each region and subregion of the
three-particle mass distribution are shown in table~\ref{table:no-3-functions}.

\begin{TABLE}{
\begin{tabular}{|c|c|c|c|}
\hline
Region&Subreg.~1&Subreg.~2&Subreg.~3\\
\hline
1 & 3 & 3 &   \\
2 & 3 & 3 &   \\
3 & 3 &   &   \\
4 & 4 & 4 &  4 \\
\hline
\end{tabular}
\caption{Number of distinct functions for the three-particle mass
$m_{cba}$.}
\label{table:no-3-functions}}
\end{TABLE}

In the three-particle case we can no longer remove $R_{C}$ from the
discussion of feet. However, other simplifications arise. One can show that
none of the four regions of the $m_{cba}$ distribution has a drop
as the last function piece goes to zero at the kinematical endpoint
for all regions and subregions. What remains is the possibility of a foot.
In figure~\ref{fig:feetcba}, the ratio $r$ is shown for two values of $R_{C}$.

\FIGURE[ht]{
\epsfxsize 14cm
\epsfbox{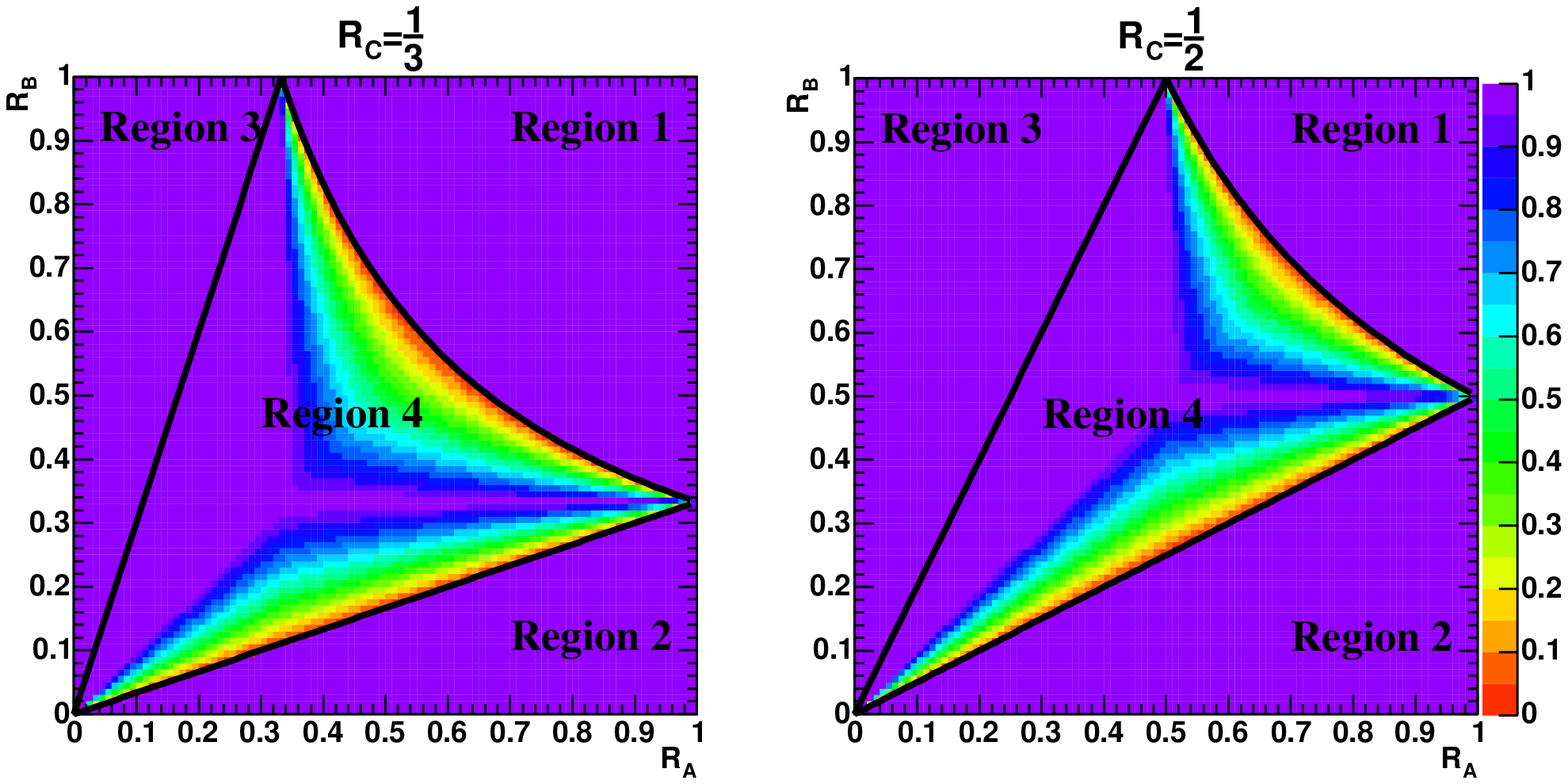}
\caption{Value of $r$ in all four regions of the $m_{cba}$ distribution,
plotted in the $R_{A}R_{B}$-plane for $R_{C} = \third$ (left)
and $R_{C} = \half$ (right).
\label{fig:feetcba}}
}

In the entirety of Regions~1, 2 and 3 we have $r=1$ with no danger of
feet because the global maximum is in the last function piece. Only in
Region~4 can a foot occur. The dangerous regions are near the border
between Region~4 and either Region~1 or Region~2, which have low
values of $r$ and potentially a misleading foot. Comparing the plots
for $R_{C} = \third$ and $R_{C} = \half$ we see how the value of $r$
changes with $R_{C}$ inside Region~4, and in particular observe that
the points where Regions~1 and 2 meet and where Regions~1 and 3 meet
also depend on $R_{C}$. However, for $m_{cba}$, the final function
piece often has a steep negative slope after the maximum, which 
reduces the possible negative effect of a foot.

\subsection{SPS benchmark points}
In figures~\ref{fig:spshigh} and \ref{fig:spslow} we show the value of $r$ in
the $m_{1/2}$--$m_0$-planes around the Snowmass mSUGRA points SPS1a, SPS1b,
SPS3 and SPS5, for the $m_{\cHigh}$ and $m_{\cLow}$ distributions
respectivly. In the top left panel we have $\tan\beta=10$ and $A_0=-m_0$ and
in the upper right panel, $\tan\beta=30$ and $A_0=0$. The lower left panel has
$\tan\beta=10$ and $A_0=0$ and the lower right panel, $\tan\beta=5$ and
$A_0=-1000$~GeV. We have only considered parameter values where we have decay
chains of the type given in eq.~(\ref{eq:squarkchain}), with an on-shell,
right-handed slepton. For the analogous decay chain via left-handed sleptons,
there is no danger of feet for values of $m_{1/2}$ up to 1~TeV. In the gray
area we have $m_{\tilde{\chi}_2^0}<m_{\tilde{l}_R}$, thus the decay is only
possible via a virtual slepton.  The white area, marked TF, is theoretically
forbidden and the light brown area has a charged LSP.

\FIGURE[ht]{
\epsfxsize 15cm
\epsfbox{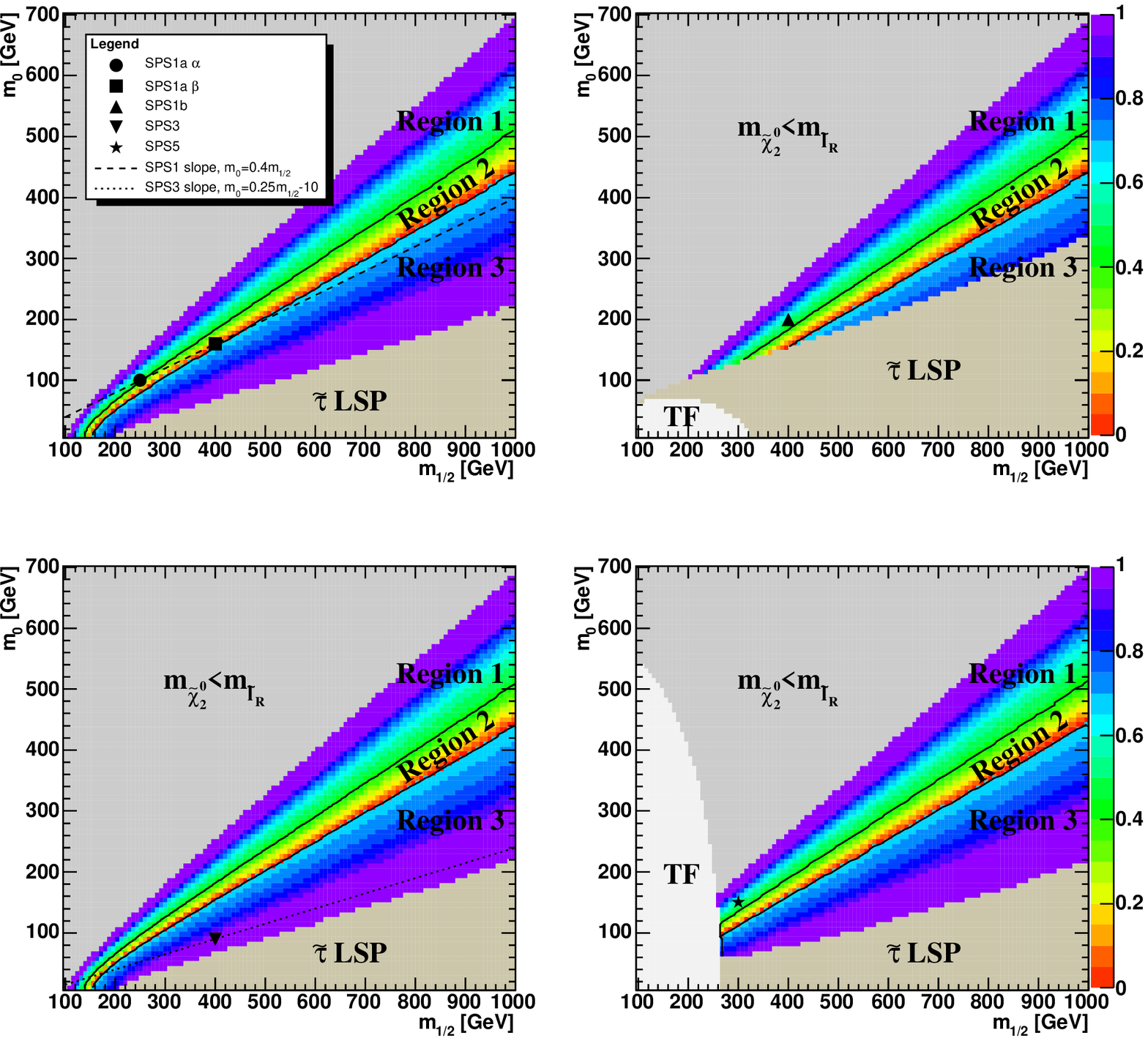}
\caption{Value of $r$ for the $m_{\cHigh}$ invariant mass distribution
in the $m_{1/2}$--$m_0$-planes around the Snowmass benchmark points
SPS1a, SPS1b, SPS3 and SPS5.
\label{fig:spshigh}}
}

\FIGURE[ht]{
\epsfxsize 15cm
\epsfbox{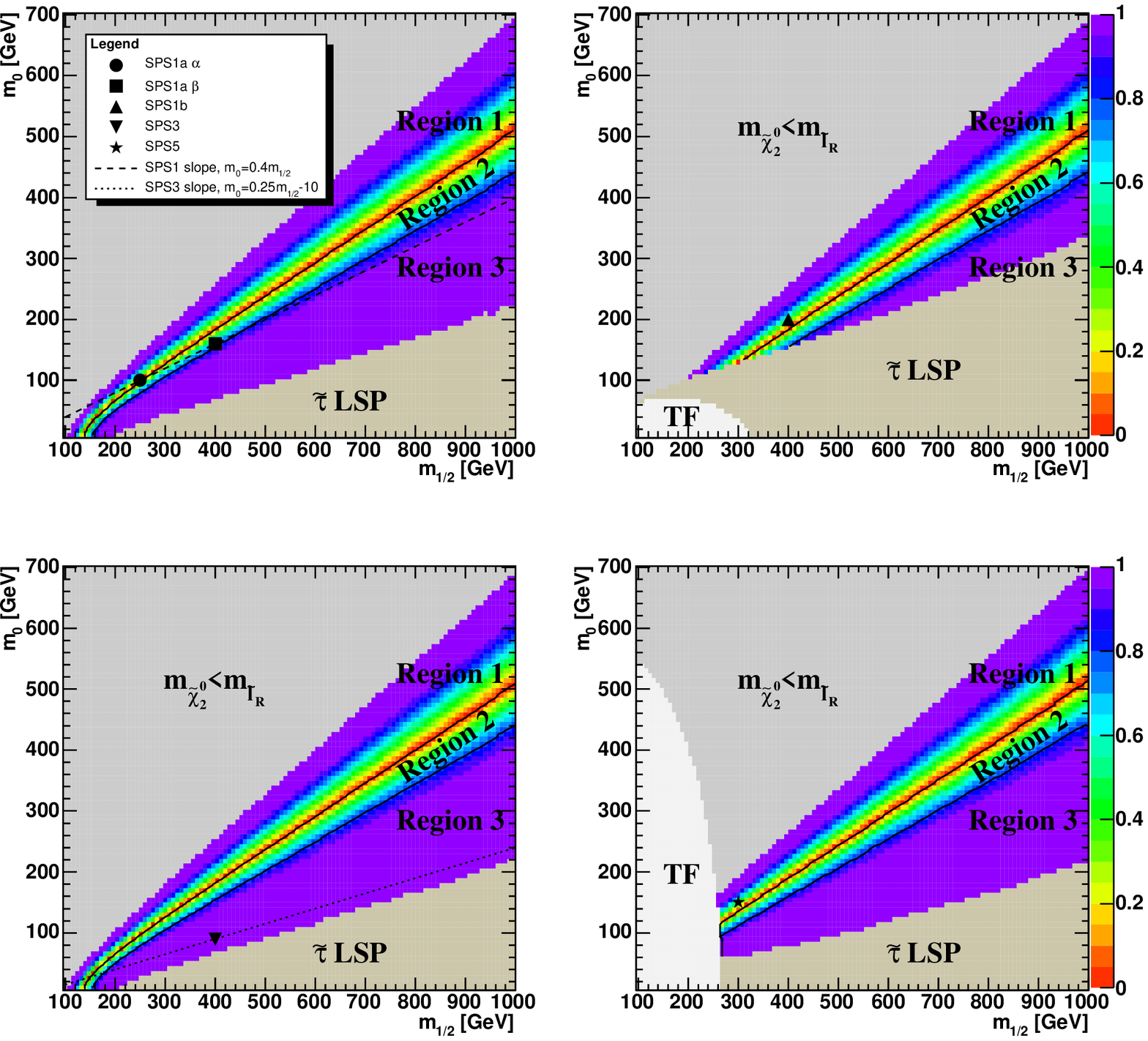}
\caption{Value of $r$ for the $m_{\cLow}$ invariant mass distribution
in the $m_{1/2}$--$m_0$-planes around the Snowmass benchmark points
SPS1a, SPS1b, SPS3 and SPS5.
\label{fig:spslow}}
}

In the mSUGRA planes we see that the dangerous area for the $m_{\cHigh}$
distribution lies in the narrow Region~2, and for the $m_{\cLow}$ distribution
in a fairly narrow corridor along the border between Regions~1 and 2. The
WMAP-consistent bulk region at low $m_{1/2}$ and $m_0$ in the top left panels
of figures~\ref{fig:spshigh} and \ref{fig:spslow}, around the point SPS1a,
lies in these dangerous areas.  This can be compared to the SPS3 parameter
line running along the WMAP-consistent stau coannihilation region in the
bottom left panel, and we see that there is little risk of problems with feet
in the stau coannihilation region for values of $m_{1/2}$ that are still
viable when we consider LEP limits on SUSY masses and on the lightest
Higgs. However, this conclusion is only valid for relatively low values of
$\tan\beta$. In the top right panel we have $\tan\beta=30$, and indeed the
dangerous area crosses the stau coannihilation region running along the border
to the region with a stau LSP. The lower right panel shows that the
WMAP-consistent stop coannihilation region, to the left of SPS5, along the
theoretically forbidden region, is also dangerous in this respect.

We do not show results for $m_{cba}$ in these parameter planes as we
find only points in Region~1 of the $m_{cba}$ distribution in the
mSUGRA planes we consider here, for values of $m_{1/2}<3$~TeV. As
discussed above, Region~1 in this distribution contains no dangerous
``feet''.

\section{Summary}
\label{sect:conclusion}
We have derived analytical expressions for the invariant mass
distributions of massless SM endproducts $c$, $b$, $a$, in cascade
decays of the form $D\rightarrow Cc\rightarrow Bbc \rightarrow
Aabc$. Our main results are valid for the decays of spin-$0$
particles, $D$, $C$, $B$, or equivalently, for a sum over all
combinations of charge and chirality in the final states. In a hadron
collider environment, the difficulty of determining the charge of
quarks from the jets they instigate, makes this a reasonable
simplification. We have discussed how different spin configurations
can easily be included in the distributions, and in the appendix show
the corresponding distributions for the specific SUSY decay chain
given in (\ref{eq:squarkchain}).

The effects of cuts, particle widths and final state radiation on the
shape of the distributions have been investigated with PYTHIA in a
particular SUSY scenario, the Snowmass benchmark point SPS1a. We find
that while a set of cuts used to remove SM background have the
potential of distorting the expected shape we can identify which cuts
are responsible for this, and in which regions of invariant mass this
distortion takes place, by looking at the distribution of the cut
parameters, and without resorting to Monte Carlo truth
information. The effect of particle widths can be compensated for by a
parametrized smearing of the distibution. Final state radiation
introduces a systematic loss of energy, thus shifting the invariant
mass distributions towards lower values.

We have also studied the effects of a generic LHC detector, in the
same SUSY model, through the use of the fast detector simulation
AcerDET. The analytic expressions for the invariant mass distributions
is seen to survive the smearing effects of the detector, and we
demonstrate a method for handling the combinatorial problem of
picking the correct jet that belongs to the decay under investigation,
from amongst the many candidates. This consistency cut method is shown
to be very effective in removing combinatorial background, but
results in a significant reduction in the number of events. The
question of whether it is optimal to use cuts to reduce this
background, or whether to try to model and subtract it, is still
open. A small, systematic shift of the distributions towards lower
invariant masses is observed. This indicates that the reconstruction
of jets and the jet recalibration routine in the detector simulation,
is insufficient for jets from this decay chain, and imply the
unsurprising conclusion that understanding the jet energy scale
will be essential in reducing systematical uncertainties in parameter
determination from the shapes of the invariant mass distributions.

We also demonstrate an application of the analytical expressions in finding
dangerous ``feet'' in the distributions, that could lead to
mismeasurements of distribution endpoints, and in turn masses. In scans
over mSUGRA parameter space we find that this danger exists for several
WMAP-consistent regions: the bulk region and the stau- and
stop-coannihilation regions.

\acknowledgments 
This research has been supported in part by the Research Council of Norway,
and the UK Particle Physics and Astronomy Research Council. PO and ARR thank
the CERN Theory Group for hospitality while part of this work was done.

\appendix

\section{Including spin in a SUSY scenario}
\label{sect:spin}
One can add spin effects in the decay chain by multiplying the integrand with
a suitable angular dependent function $f_u(u)$ or $f_v(v)$, depending on the
spin configuration in question. We will here give the results of including
spin in the SUSY decay chain (\ref{eq:squarkchain}), as discussed in the
Introduction. In this decay chain we have $A=\tilde{\chi}^{0}_{1}$,
$B=\tilde{l}$, $C=\tilde{\chi}^{0}_{2}$ and $D=\tilde{q}_{L}$. The
spin-$\half$ of $\tilde{\chi}^{0}_{2}$ will yield an extra factor of either
$2u$ or $2(1-u)$ depending on the different combinations of chirality and
charge in the final state for the quark and the ``near'' lepton, as given in
table~\ref{table:spin}. Only the decays starting from a left handed squark are
listed fully, those for the right handed squark follow from a simple
interchange of left and right handedness\footnote{Note that the handedness of
sparticles refer to the chirality of the SM particles they couple to.}.  The
``far'' lepton does not contribute any spin-dependent factor, since the LSP,
associated with that fermion line, is not observed.

\begin{TABLE}{
\begin{tabular}{|l|c|c|c|c|}
\hline
Process &chirality of $q$ &chirality of $\lN$ &chirality of $\lF$ &Factor \\
\hline
$\tilde q_L\to q\lN^-\lF^+\tilde\chi_1^0$ & L & L & L & $2u$ \\
$\tilde q_L\to q\lN^+\lF^-\tilde\chi_1^0$ & L & L & L & $2(1-u)$ \\
$\tilde{\bar{q}}_L\to \bar q\lN^+\lF^-\tilde\chi_1^0$ & L & L & L & $2u$ \\
$\tilde{\bar{q}}_L\to \bar q\lN^-\lF^+\tilde\chi_1^0$ & L & L & L & $2(1-u)$ \\
$\tilde q_L\to q\lN^-\lF^+\tilde\chi_1^0$ & L & R & R & $2(1-u)$ \\
$\tilde q_L\to q\lN^+\lF^-\tilde\chi_1^0$ & L & R & R & $2u$ \\
$\tilde{\bar{q}}_L\to \bar q\lN^+\lF^-\tilde\chi_1^0$ & L & R & R & $2(1-u)$ \\
$\tilde{\bar{q}}_L\to \bar q\lN^-\lF^+\tilde\chi_1^0$ & L & R & R & $2u$ \\
$\tilde q_R\to q\lN^-\lF^+\tilde\chi_1^0$ & R & L & L & $2(1-u)$ \\
\vdots &\vdots & \vdots & \vdots & \vdots \\
$\tilde q_R\to q\lN^-\lF^+\tilde\chi_1^0$ & R & R & R & $2u$ \\
\vdots &\vdots & \vdots & \vdots & \vdots \\
\hline
\end{tabular}
\caption{Spin factors. The chirality L/R of $l^+$ ($\bar q$) denotes
that it is the antiparticle of a left-/ right-handed $l^-$ ($q$).}
\label{table:spin}}
\end{TABLE}

Except for the first case, discussed in section~\ref{sect:m_ca},
we shall not give explicit formulas for both chirality cases.
The two cases are related as follows:
\begin{equation}
\frac{1}{\Gamma} 
\frac{\partial \Gamma}{\partial m^2}\bigg|_{2(1-u)}
=\frac{2}{\Gamma_0} \frac{\partial \Gamma_0}{\partial m^2} 
-\frac{1}{\Gamma} 
\frac{\partial \Gamma}{\partial m^2}\bigg|_{2u}.
\end{equation}

The integrated decay widths $\Gamma$ are for both chirality configurations
equal to $\Gamma_0$, since the spin correlations only introduce
forward-backward asymmetries (in the $\tilde \chi_2^0$ rest frame), which
integrate to zero.
\subsection{$m_{ca}$} \label{sect:m_ca}
We begin by including spin in the $m_{ca}$ distribution by multiplying
the RHS of eq.~(\ref{eq:d2Gdudv}) with $f_u(u)$. We find for the
$f_u(u)=2u$ case that
\begin{equation}
\frac{1}{\Gamma} \frac{\partial \Gamma}{\partial m_{ca}^2} 
= \left\{ \begin{array}{lcl}
\displaystyle
\frac{2}{(m_{ca}^{\max})^{2}a^2}
\left[\ln\frac{m_{C}^2}{m_{B}^2}-a\right]
&\; {\rm for }\;&  
\displaystyle 0<m_{ca}<\frac{m_{B}}{m_{C}} m_{ca}^{\max},\\[5mm]
\displaystyle
\frac{2}{(m_{ca}^{\max})^{2}a^2}
\left[\ln\frac{(m_{ca}^{\max})^2}{m_{ca}^2}
+\frac{m_{ca}^2}{(m_{ca}^{\max})^2}-1\right]
&\; {\rm for }\;&
\displaystyle \frac{m_{B}}{m_{C}} m_{ca}^{\max}
<m_{ca}<m_{ca}^{\max},
\end{array} \right.
\end{equation}
and for $f_u(u)=2(1-u)$ we have
\begin{equation}
\frac{1}{\Gamma} \frac{\partial \Gamma}{\partial m_{ca}^2} 
= \left\{ \begin{array}{lcl}
\displaystyle
\frac{2}{(m_{ca}^{\max})^{2}a^2}
\left[\frac{m_B^2}{m_C^2}\ln\frac{m_{B}^2}{m_{C}^2}+a\right]
&\; {\rm for }\;&  
\displaystyle 0<m_{ca}<\frac{m_{B}}{m_{C}} m_{ca}^{\max},\\[5mm]
\displaystyle
\frac{2}{(m_{ca}^{\max})^{2}a^2}
\left[\frac{m_B^2}{m_C^2}\ln\frac{m_{ca}^2}{(m_{ca}^{\max})^2}
-\frac{m_{ca}^2}{(m_{ca}^{\max})^2}+1\right]
&\; {\rm for }\;&  
\displaystyle \frac{m_{B}}{m_{C}} m_{ca}^{\max}
<m_{ca}<m_{ca}^{\max}.
\end{array} \right.
\end{equation}
This confirms the result of \cite{Smillie:2005ar}.
As required, the average of the two again gives eq.~(\ref{eq:mca}).

\subsection{$\mcHigh$}
For Region~1 of the $\mcHigh$ distribution, multiplying (\ref{eq:d2Gdxdmhigh})
by $f_u(u)=2u$ gives
\begin{equation}
\frac{1}{\Gamma} \frac{\partial \Gamma}{\partial m_\cHigh^2} = \left\{
\begin{array}{lcl} \displaystyle
\lefteqn{\frac{2}{(m_{ca}^{\max})^{2}a^2}
\left[
\ln\frac{(m_{cb}^{\max})^{2}}{(m_{cb}^{\max})^{2}-am_\cHigh^{2}}
-\frac{am_\cHigh^{2}}{(m_{cb}^{\max})^{2}}
\frac{(m_{cb}^{\max})^{2}-2am_\cHigh^{2}}{(m_{cb}^{\max})^{2}-am_\cHigh^{2}}
\right]} && \\[6mm]
&\text{for~~}& 0<m_\cHigh<m_{cb}^{\max}, \\[5mm] \displaystyle
\frac{2}{(m_{ca}^{\max})^{2}a^2}
\left[\ln\frac{m_{C}^{2}}{m_{B}^{2}}-a\right] \hspace{1cm} &\text{for~~}&
m_{cb}^{\max}<m_\cHigh<\frac{m_{B}}{m_{C}}
m_{ca}^{\max}, \\[5mm] \displaystyle
\lefteqn{\frac{2}{(m_{ca}^{\max})^{2}a^2}
\left[
\ln\frac{(m_{ca}^{\max})^{2}}{m_\cHigh^{2}}
+\frac{m_\cHigh^{2}}{(m_{ca}^{\max})^{2}}-1
\right]} && \\[6mm]
 &\text{for~~} & \frac{m_{B}}{m_{C}}m_{ca}^{\max}<m_\cHigh<m_{ca}^{\max}.
\end{array} \right. 
\end{equation}
For Region~2 we get
\begin{equation}
\frac{1}{\Gamma} \frac{\partial \Gamma}{\partial m_\cHigh^2} = \left\{
\begin{array}{lcl} \displaystyle
\lefteqn{\frac{2}{(m_{ca}^{\max})^{2}a^2}
\left[
\ln\frac{(m_{cb}^{\max})^{2}}{(m_{cb}^{\max})^{2}-am_{\cHigh}^{2}}
-\frac{am_{\cHigh}^{2}}{(m_{cb}^{\max})^{2}}
\frac{(m_{cb}^{\max})^{2}-2am_\cHigh^{2}}{(m_{cb}^{\max})^{2}-am_\cHigh^{2}}
\right]} \hspace{60mm}  &&\\[5mm]
&\text{for~~}& 
0<m_{\cHigh}<m_{c2{\rm (eq)}}^{\max},
\\[5mm] \displaystyle
\lefteqn{\frac{2}{(m_{ca}^{\max})^{2}a^2}
\left[
\ln\frac{(m_{ca}^{\max})^{2}}{m_{\cHigh}^{2}}
+\frac{a^2(m_{ca}^{\max})^{2}m_{\cHigh}^{2}}{(m_{cb}^{\max})^{4}}
+\frac{m_{\cHigh}^{2}}{(m_{ca}^{\max})^{2}}-1
\right]} &&\\[5mm]
&\text{for~~}&
m_{c2{\rm (eq)}}^{\max}<m_{\cHigh}<m_{cb}^{\max},
\\[5mm] \displaystyle
\frac{2}{(m_{ca}^{\max})^{2}a^2}
\left[
\ln\frac{(m_{ca}^{\max})^{2}}{m_{\cHigh}^{2}}
+\frac{m_{\cHigh}^{2}}{(m_{ca}^{\max})^{2}}-1
\right]
&\text{for~~} &
m_{cb}^{\max}<m_{\cHigh}<m_{ca}^{\max}.
\end{array} \right. 
\end{equation}
For Region~3 we have
\begin{equation}
\frac{1}{\Gamma} \frac{\partial \Gamma}{\partial m_\cHigh^2} = \left\{
\begin{array}{lcl} \displaystyle
\lefteqn{\frac{2}{(m_{ca}^{\max})^{2}a^2}
\left[
\ln\frac{(m_{cb}^{\max})^{2}}{(m_{cb}^{\max})^{2}-am_{\cHigh}^{2}}
-\frac{am_{\cHigh}^{2}}{(m_{cb}^{\max})^{2}}
\frac{(m_{cb}^{\max})^{2}-2am_\cHigh^{2}}{(m_{cb}^{\max})^{2}-am_\cHigh^{2}}
\right]} \hspace{60mm}  &&\\[6mm]
&\text{for~~}& 
0<m_{\cHigh}<m_{c2{\rm (eq)}}^{\max},
\\[5mm] \displaystyle
\lefteqn{\frac{2}{(m_{ca}^{\max})^{2}a^2}
\left[
\ln\frac{(m_{ca}^{\max})^{2}}{m_{\cHigh}^{2}}
+\frac{a^2(m_{ca}^{\max})^{2}m_{\cHigh}^{2}}{(m_{cb}^{\max})^{4}}
+\frac{m_{\cHigh}^{2}}{(m_{ca}^{\max})^{2}}-1
\right]} &&\\[6mm]
&\text{for~~}&
m_{c2{\rm (eq)}}^{\max}<m_{\cHigh}<m_{ca}^{\max},
\\[5mm] \displaystyle
\frac{2m_{\cHigh}^{2}}{(m_{cb}^{\max})^{4}}
&\text{for~~} &
m_{ca}^{\max}<m_{\cHigh}<m_{cb}^{\max}.
\end{array} \right. 
\end{equation}

\subsection{$\mcLow$}
For the $\mcLow$ distribution, in Region~1, and with $f_u(u)=2u$, we get
\begin{align}
& \frac{1}{\Gamma} \frac{\partial \Gamma}{\partial m_\cLow^2} =
\frac{2}{(m_{ca}^{\max})^{2}a^2}
\left[
\ln\frac{(m_{cb}^{\max})^{2}-am_\cLow^{2}}
{\frac{m_B^2}{m_C^2}(m_{cb}^{\max})^{2}}
+\frac{a^2m_\cLow^{2}(m_{ca}^{\max})^{2}}{(m_{cb}^{\max})^{4}}
\right.
\nonumber \\[5mm]
& \left. \hspace{4.5cm}
-\frac{a^2m_\cLow^{4}}{(m_{cb}^{\max})^{2}
\left[(m_{cb}^{\max})^{2}-am_\cLow^{2}\right]}
+\frac{am_\cLow^{2}}{(m_{cb}^{\max})^{2}}-a
\right],
\nonumber \\[2mm]
\end{align}
for $0<m_\cLow<m_{cb}^{\max}$.
For Regions~2 and 3 the distribution is for $f_u(u)=2u$ given by
\begin{align}
&\frac{1}{\Gamma} \frac{\partial \Gamma}{\partial m_\cLow^2}
=
\nonumber \\[6mm]
&\frac{2}{(m_{ca}^{\max})^{2}a^2}\left\{
\begin{array}{l} \displaystyle
\lefteqn{
\ln\frac{(m_{cb}^{\max})^2-am_\cLow^2}{\frac{m_B^2}{m_C^2}(m_{cb}^{\max})^2}
+\frac{a^2(m_{ca}^{\max})^{2}m_\cLow^2}{(m_{cb}^{\max})^{4}}}
\\[6mm] \displaystyle \hspace{10mm}
-\frac{a^2m_\cLow^4}{(m_{cb}^{\max})^2
\left[(m_{cb}^{\max})^2-am_\cLow^2\right]}
+\frac{am_\cLow^2}{(m_{cb}^{\max})^{2}}-a
\\[8mm] \hspace{20mm}
\text{for~~}
\displaystyle 0<m_{\cLow}<\frac{m_{B}}{m_{C}}m_{ca}^{\max},
\\[6mm] \displaystyle
\lefteqn{
\ln\frac{(m_{ca}^{\max})^{2}\left[(m_{cb}^{\max})^{2}-am_\cLow^2\right]}
{(m_{cb}^{\max})^{2}m_\cLow^2}
+\frac{a^2(m_{ca}^{\max})^{2}m_\cLow^2}{(m_{cb}^{\max})^{4}}}
\\[6mm] \displaystyle \hspace{10mm}
-\frac{a^2m_\cLow^4}{(m_{cb}^{\max})^2
\left[(m_{cb}^{\max})^2-am_\cLow^2\right]}
+\frac{m_\cLow^2}{(m_{ca}^{\max})^{2}}+\frac{am_\cLow^2}{(m_{cb}^{\max})^2}-1
\\[8mm] \hspace{20mm}
\text{for~~}
\displaystyle \frac{m_B}{m_C}m_{ca}^{\max}< m_\cLow<m_{c2{\rm (eq)}}^{\max}. 
\end{array} \right. \nonumber
\\[2mm]
\end{align}

\subsection{$m_{cba}$}
In the $m_{cba}$ distribution the spin factor enters into the
integral $L(a_1,a_2)$ of eq.~(\ref{eq:define-L}).
We define a new integral:
\begin{eqnarray}
M(a_1,a_2) & \equiv & 
\int_{a_{1}}^{a_{2}}
\frac{y}{\sqrt{y^{2}+2ym_{B}+m_{B}^{2}-m_{D}^{2}}}dy \nonumber \\
& = & \left.\sqrt{y^2+2ym_B+m_B^2-m_D^2}\ \right|^{a_2}_{a_1}-m_BL(a_1,a_2).
\label{eq:define-M}
\end{eqnarray}
For all four Regions of the $m_{cba}$ distribution the effect of the
neutralino spin can be added by making the following simple
substitution in the distributions of
eqs.~(\ref{eq:region11})--(\ref{eq:region43}), for the $f_u(u)=2u$
case:
\begin{equation}
L(a_1,a_2)\rightarrow L'(a_1,a_2)=
\frac{2(m_D^2-m_B^2)m_C^2L(a_1,a_2)-4m_Bm_C^2M(a_1,a_2)}
{(m_D^2-m_C^2)(m_C^2-m_B^2)}.
\end{equation}



\begin{thebibliography}{100}

\bibitem{Wess:1974tw}
J.~Wess and B.~Zumino,
Nucl.\ Phys.\ B {\bf 70}, 39 (1974).

\bibitem{Fayet:1976cr}
P.~Fayet and S.~Ferrara,
Phys.\ Rept.\  {\bf 32} (1977) 249.

\bibitem{Dimopoulos:1981zb}
S.~Dimopoulos and H.~Georgi,
Nucl.\ Phys.\ B {\bf 193} (1981) 150.

\bibitem{Nilles:1983ge}
H.~P.~Nilles,
Phys.\ Rept.\  {\bf 110} (1984) 1.

\bibitem{Haber:1984rc}
H.~E.~Haber and G.~L.~Kane,
Phys.\ Rept.\  {\bf 117} (1985) 75.

\bibitem{Hinchliffe:1996iu}
I.~Hinchliffe, F.~E.~Paige, M.~D.~Shapiro, J.~Soderqvist and W.~Yao,
Phys.\ Rev.\ D {\bf 55} (1997) 5520
[arXiv:hep-ph/9610544].

\bibitem{Hinchliffe:1998zj}
I.~Hinchliffe, F.~E.~Paige, E.~Nagy, M.~D.~Shapiro, J.~Soderqvist and W.~Yao,
LBNL-40954

\bibitem{Bachacou:1999zb}
H.~Bachacou, I.~Hinchliffe and F.~E.~Paige,
Phys.\ Rev.\ D {\bf 62} (2000) 015009
[arXiv:hep-ph/9907518].

\bibitem{Polesello}
G. Polesello,
{\it Precision SUSY measurements with
ATLAS for SUGRA point 5},
ATLAS Internal Note, PHYS-No-111, October 1997. 

\bibitem{Allanach:2000kt}
B.~C.~Allanach, C.~G.~Lester, M.~A.~Parker and B.~R.~Webber,
JHEP {\bf 0009} (2000) 004
[arXiv:hep-ph/0007009].

\bibitem{Lester} C. G. Lester, 
{\it Model independent sparticle mass measurements at ATLAS},
Ph.~D.\ thesis,
http://www.slac.stanford.edu/spires/find/hep/www?r=cern-thesis-2004-003

\bibitem{Gjelsten:2004}
B.~K.~Gjelsten, D.~J.~Miller and P.~Osland,
JHEP {\bf 12} (2004) 003
[arXiv:hep-ph/0410303].

\bibitem{Gjelsten:2005aw}
B.~K.~Gjelsten, D.~J.~Miller and P.~Osland,
JHEP {\bf 06} (2005) 015
[arXiv:hep-ph/0501033].

\bibitem{Allanach:2002nj}
B.~C.~Allanach {\it et al.},
in {\it Proc. of the APS/DPF/DPB Summer Study on the Future of
Particle Physics (Snowmass 2001) } ed. N.~Graf,
Eur.\ Phys.\ J.\ C {\bf 25} (2002) 113
[eConf {\bf C010630} (2001) P125]
[arXiv:hep-ph/0202233].

\bibitem{Appelquist:2000nn}
T.~Appelquist, H.~C.~Cheng and B.~A.~Dobrescu,
Phys.\ Rev.\ D {\bf 64}, 035002 (2001)
[arXiv:hep-ph/0012100].

\bibitem{Cheng:2002ab}
H.~C.~Cheng, K.~T.~Matchev and M.~Schmaltz,
Phys.\ Rev.\ D {\bf 66} (2002) 056006
[arXiv:hep-ph/0205314].

\bibitem{PYTHIA}
T. Sj\"ostrand, P. Ed\'en, C. Friberg, L. L\"onnblad, G. Miu, 
S. Mrenna, E. Norrbin, Comput.\ Phys.\ Commun. 135 (2001) 238;
T.~Sj\"ostrand, L.~L\"onnblad and S.~Mrenna,
``PYTHIA 6.2: Physics and manual'',
arXiv:hep-ph/0108264.

\bibitem{Richardson:2001df}
P.~Richardson,
JHEP {\bf 0111} (2001) 029
[arXiv:hep-ph/0110108].

\bibitem{Barr:2004ze}
A.~J.~Barr,
arXiv:hep-ph/0405052.

\bibitem{Smillie:2005ar}
J.~M.~Smillie and B.~R.~Webber,
arXiv:hep-ph/0507170.

\bibitem{Birkedal:2005cm}
A.~Birkedal, R.~C.~Group and K.~Matchev,
arXiv:hep-ph/0507002.

\bibitem{Baer:1993ae}
H.~Baer, F.~E.~Paige, S.~D.~Protopopescu and X.~Tata,
arXiv:hep-ph/9305342,
arXiv:hep-ph/0001086.

\bibitem{Lai:1999wy}
H.~L.~Lai {\it et al.}  [CTEQ Collaboration],
Eur.\ Phys.\ J.\ C {\bf 12} (2000) 375
[arXiv:hep-ph/9903282].

\bibitem{Richter-Was:2002ch}
E.~Richter-Was,
arXiv:hep-ph/0207355.

\bibitem{ATLFAST}
E.~Richter-Was, D.~Froidevaux and L.~Poggioli,
``ATLFAST 2.0: a fast simulation package for ATLAS'',
Tech.\ Rep.\ ATL-PHYS-98-131 (1998).

\end{thebibliography}
\end{document}